\documentclass[aip,jcp,superscriptaddress,amsmath,amssymb,twocolumn,reprint]{revtex4-1}

\usepackage[version=3]{mhchem} 
\usepackage{graphicx}
\usepackage{amsmath}
\usepackage{amssymb}
\usepackage{nicefrac}
\usepackage{booktabs}

\usepackage{array}
\newcolumntype{m}{>{$} c <{$}}
\usepackage{color}

\usepackage{braket}
\usepackage{enumitem} 
\usepackage{adjustbox}
\usepackage{lipsum}
\newcommand{\tagarray}{%
\mbox{}\refstepcounter{equation}%
$(\theequation)$%
}

\def\fv{{\bf f}}

\def\beq{\begin{equation}}
\def\eeq{\end{equation}}

\def\br{{\bf r}}
\def\bs{{\bf s}}
\def\bx{{\bf x}}
\def\fv{{\bf f}}

\def\md{\mathrm{d}}
\begin{document} 

\author{Sara Giarrusso}
\affiliation
{Department of Theoretical Chemistry and Amsterdam Center for Multiscale Modeling, FEW, Vrije Universiteit, De Boelelaan 1083, 1081HV Amsterdam, The Netherlands}
\author{Stefan Vuckovic}
\affiliation
{Department of Theoretical Chemistry and Amsterdam Center for Multiscale Modeling, FEW, Vrije Universiteit, De Boelelaan 1083, 1081HV Amsterdam, The Netherlands}
\author{Paola Gori-Giorgi}
\affiliation
{Department of Theoretical Chemistry and Amsterdam Center for Multiscale Modeling, FEW, Vrije Universiteit, De Boelelaan 1083, 1081HV Amsterdam, The Netherlands}
\email{p.gorigiorgi@vu.nl}

\title{Response potential in the strong-interaction limit of DFT: Analysis and comparison with the coupling-constant average}
\begin{abstract}
Using the formalism of the conditional amplitude, we study the response part of the exchange-correlation potential in the strong-coupling limit of density functional theory, analysing its peculiar features and comparing it with the response potential averaged over the coupling constant for small atoms and for the hydrogen molecule.
We also use a simple one-dimensional model of a stretched heteronuclear molecule to derive exact properties of the response potential in the strong-coupling limit. The simplicity of the model allows us to unveil relevant features also of the exact Kohn-Sham potential and its different components, namely the appearance of a second peak in the correlation kinetic potential on the side of the most electronegative atom.
\end{abstract}
\maketitle

\section{Introduction}
Kohn-Sham (KS) Density Functional Theory (DFT) \cite{KS65} is the most used tool in quantum chemistry calculations thanks to its ability to predict properties of interest of a variety of physical, chemical and biochemical systems at an acceptable computational cost with a reasonable accuracy. 
Nonetheless, there are still many relevant cases, typically when electron correlation plays a prominent role, in which current KS DFT methodologies are deficient, making the quest for new approximations to the unknown piece of information in DFT, the so-called exchange-correlation (XC) functional, an active research field.\cite{CohMorYan-CR-12, SuXu-ARPC-17, MarHea-MP-17}
This quest for a better (more versatile or accurate) and at the same time computationally affordable XC functional cannot proceed without a synchronized understanding of its exact properties and a constant search to find new ones that can act as constraints to build approximations.\cite{LevPer-PRA-85, LevPer-PRB-93, PerStaTaoScu-PRA-08, PerRuzTaoStaScuCso-JCP-05}
In this context, a very important role is played by studies\cite{BuiBaeSni-PRA-89,  UmrGon-PRA-94, GriLeeBae-JCP-94, FilGonUmr-INC-96, BaeGri-PRA-96, GriLeeBae-JCP-96, BaeGri-JPCA-97, TemMarMai-JCTC-09, HelTokRub-JCP-09, CueAyeSta-JCP-15, CueSta-MP-16, KohPolSta-PCCP-16, HodRamGod-PRB-16, YinBroLopVarGorLor-PRB-16, BenPro-PRA-16, RyaOspSta-JCP-17, HodKraSchGro-JPCL-17} focussing on the XC \emph{potential} given by the functional derivative of the XC functional, $v_{xc}(\mathbf{r})=\frac{\delta E_{xc}[\rho]}{\delta\rho(\br)}$, whose properties are crucial, for example, to accurately predict static electric polarizabilities and band gaps, and to correctly describe strongly-correlated systems and bond breaking.

Pioneering work in this direction was pursued by Baerends and coworkers, who have analysed the XC potential, deriving exact expressions in terms of wavefunctions and KS quantities.\cite{BuiBaeSni-PRA-89, GriLeeBae-JCP-94, LeeGriBae-ZPD-95, BaeGri-PRA-96, GriLeeBae-JCP-96, BaeGri-JPCA-97}
Their work builds on the theory of conditional probability amplitudes first developed by Hunter,\cite{Hun-IJQC-75-1, Hun-IJQC-75-2} which yields an exact differential equation for the square root of the density,\cite{Hun-IJQC-75-2} and was introduced in a DFT context by Levy, Perdew and Sahni.\cite{LevPerSah-PRA-84} Baerends and coworkers have applied the same formalism to the KS hamiltonian, deriving an insightful and exact decomposition of the XC potential, into so-called kinetic, response, and XC hole terms,\cite{BuiBaeSni-PRA-89, GriLeeBae-JCP-94, LeeGriBae-ZPD-95, BaeGri-PRA-96, GriLeeBae-JCP-96, BaeGri-JPCA-97} showing that each contribution has different properties and peculiarities that should be approximated with different standards.\cite{GriLeeLenBae-PRA-95, LeeGriBae-ZPD-95, KuiOjaEnkRan-PRB-10, GriMenBae-JCP-16} For example, the response part builds a step structure in the KS potential of a stretched heteronuclear molecule, and the kinetic part builds a peak in the midbond region of a stretched bond.\cite{BaeGri-JPCA-97}
Lately this subject has  gained renovated interest for various reasons, spanning from the construction of KS potentials from wavefunctions in small basis sets,\cite{CueAyeSta-JCP-15, CueSta-MP-16, KohPolSta-PCCP-16, RyaOspSta-JCP-17} to the use of response potential approximations to compute band gaps\cite{KuiOjaEnkRan-PRB-10} and correcting semilocal functionals,\cite{GriMenBae-JCP-16} to further investigations of the step structure for molecular dissociation,\cite{TemMarMai-JCTC-09, HodKraSchGro-JPCL-17} and of the kinetic peak for Mott insulators.\cite{HelTokRub-JCP-09, YinBroLopVarGorLor-PRB-16}

At the same time, in recent years,  the mathematical structure of the limit of infinite interaction strength in DFT, corresponding to the so-called strictly-correlated electrons (SCE) functional, has been thoroughly investigated.\cite{SeiGorSav-PRA-07, GorVigSei-JCTC-09, ButDepGor-PRA-12, CotFriKlu-CPAM-13, Lew-arxiv-17, CotFriKlu-arxiv-17} The SCE functional has a highly non-local density dependence, but its functional derivative can be computed via a physically transparent and rigorous auxiliary equation, which provides a powerful shortcut to access the corresponding XC potential.\cite{MalGor-PRL-12} This SCE XC potential has been used in the KS framework to compute properties of electrons confined at low-density, close to the ``Wigner-molecular" regime.\cite{MalGor-PRL-12, MalMirCreReiGor-PRB-13, MirSeiGor-PRL-13, MenMalGor-PRB-14} Despite how extreme the SCE limit might sound, it has the advantage of unveiling explicitly how the density is transformed into an electron-electron interaction, in a well defined asymptotic case (low-density or strong interaction) for the exact XC functional.\cite{Lew-arxiv-17, CotFriKlu-arxiv-17} Its peculiar mathematical structure has already inspired new approximations, in which instead of the traditional DFT ingredients (local density, density gradients, KS orbitals), certain integrals of the density play a crucial role.\cite{WagGor-PRA-14, BahZhoErn-JCP-16, VucGor-JPCL-17} 

The SCE limit, however, has never been analyzed from the point of view of the conditional amplitude framework, and nothing is known about the behavior of the different components of the corresponding XC potential. It is the main purpose of this work to fill this gap. We start by generalising the Schr\"odinger equation for the square root of the density to any coupling-strength $\lambda$ value, analysing its features in the $\lambda\to\infty$ (or SCE) limit (sec.~\ref{Ext-lambda}). We derive and analyse the response part of the SCE exchange-correlation potential (sec.~\ref{sec:vrespvarious}), and compare it with the one from the coupling-constant average formalism for small atoms and the H$_2$ molecule (sec.~\ref{physical}). Using a one-dimensional model\cite{TemMarMai-JCTC-09, HelTokRub-JCP-09, HodKraSchGro-JPCL-17} for the dissociation of a heteroatomic molecule, we analyze in this limit the SCE and exact exchange-correlation potentials, focusing on the step structure and further analyzing the kinetic potential for the physical coupling strength (sec.~\ref{sec:ABmodel}).

\section{Strong-interaction limit of the effective equation for the square root of the density} \label{Ext-lambda}

Consider the $\lambda$-dependent Hohenberg--Kohn functional within the constrained-search definition \cite{Lev-PNAS-79}
\begin{equation}
F_{\lambda}[\rho]= \min_{\Psi\to\rho}\langle \Psi | \hat{T} + \lambda \hat{V}_{ee} | \Psi \rangle,
\label{eq:Flambda}
\end{equation}
 assuming that $\rho$ is \textit{v}-representable for all $\lambda$, one can write a series of $\lambda$-dependent hamiltonian with  fixed density
 \begin{equation}
\hat{H}_{\lambda}= \hat{T} + \lambda \hat{V}_{ee} + \hat{V}^{\lambda},
\label{AC}
\end{equation}
where $\hat{V}^{\lambda}=\sum_i^N v^{\lambda}(\br_i)$ and
\begin{equation}\label{eq:vextlambda}
v ^\lambda [\rho] (\br) = -\frac{\delta F_{\lambda}[\rho]}{\delta \rho (\br)}
\end{equation}
is the local external potential that delivers the prescribed density as the ground-state density of hamiltonian \ref{AC} at each $\lambda$, i.e. $\rho_\lambda(\br)=\rho_1(\br)\equiv \rho(\br)$, and $\Psi_{\lambda}(1,..., N)$ -- with $1,...,N$ the spin-spatial coordinates of the $N$ electrons -- is the ground state wavefunction of hamiltonian \ref{AC} at each $\lambda$.
Following refs~\citenum{Hun-IJQC-75-1, Hun-IJQC-75-2, LevPerSah-PRA-84, BuiBaeSni-PRA-89}, we partition the hamiltonian \ref{AC} as
\begin{equation} \label{PH}
\hat{H}_{\lambda}^N =- \frac{\nabla^2_1}{2}+\lambda \sum_{p>1}\frac{1}{r_{1p}} +  v^{\lambda}(\br_1)+\hat{H}_{\lambda*}^{N-1} 
\end{equation}
where $\hat{H}_{\lambda*}^{N-1}$ is the Hamiltonian \ref{AC} deprived of one particle (which is in general different from the $\lambda$-dependent Hamiltonian of the physical $(N-1)$-electron system),
and factorize the wavefunction $\Psi_{\lambda}(1,..., N)$ as
\begin{equation} \label{FWF}
\Psi_{\lambda}(1,..., N) = \sqrt{\frac{\rho(\br)}{N}} \Phi_{\lambda}(\sigma,2,...,N|\br),
\end{equation}
with $1= \br \sigma$ being the spatial-spin coordinates of electron $1$ taken as a reference. The function $\Phi_{\lambda}(\sigma,2,...,N|\br)$ is called {\it conditional amplitude} and describes the behavior of the remaining $N-1$ electrons as a parametric function of the position $\br$ of electron 1. Notice that the conditional amplitude, when integrated over its $N-1$ variables, is normalized to 1 for all values of $\br$.

By applying eq~\ref{PH} to eq~\ref{FWF}, multiplying by $\Phi^*_{\lambda}(\sigma,2,...,N|\br)$ to the left and integrating over all variables except $\br$ as in refs~\citenum{Hun-IJQC-75-1, Hun-IJQC-75-2, LevPerSah-PRA-84, BuiBaeSni-PRA-89}, one obtains an effective equation for the square root of the density for any $ \lambda$-value,
\begin{equation} \label{ELE}
\left(\!\!-\,\frac{\nabla^2}{2}+v_{\lambda,\, eff}(\br)+v^{\lambda}(\br)\!\right)\!\!\sqrt{\rho(\br)}\!=\!(E_{0,\lambda}^{N}-E_{0,\lambda*}^{N-1})\!\sqrt{\rho(\br)},
\end{equation}
where
 \begin{equation}
v_{\lambda, \, eff}(\br)=   v_{\lambda, N-1}(\br)+v_{\lambda, kin}(\br)+\lambda \, v_{\lambda, cond}(\br),
\end{equation} 
and $E_{0,\lambda*}^{N-1}$ is the ground-state energy of the $N-1$ system in the same effective potential as the $N$-particle one (thus $E_{0,\lambda*}^{N-1} = E_{0,\lambda}^{N-1}$ only for $\lambda =1$). The various components of the effective potential have each its own physical meaning and peculiar features, and have been carefully studied by many authors.\cite{BuiBaeSni-PRA-89, GriLeeBae-JCP-94, LeeGriBae-ZPD-95, BaeGri-PRA-96, GriLeeBae-JCP-96, BaeGri-JPCA-97, KohPolSta-PCCP-16,TemMarMai-JCTC-09, HodKraSchGro-JPCL-17, HelTokRub-JCP-09, YinBroLopVarGorLor-PRB-16} 
The term $v_{\lambda, N-1}(\br)$ is related to the response potential (see sec.~\ref{sec:vrespvarious} below) and is given by
\begin{align}\label{N-1}
 & v_{\lambda, N-1}(\br)= \nonumber \\
&\int \Phi^*_{\lambda}(\sigma,2,...|\br) \hat{H}_{\lambda*}^{N-1}\Phi_{\lambda}(\sigma,2,...|\br) \md\sigma\md2...\md N- E_{0,\lambda*}^{N-1},
\end{align}
where the subtraction of the quantity $E_{0,\lambda*}^{N-1}$ makes this potential go to zero when $|\br|\to\infty$, as in this case the conditional amplitude usually collapses to the ground state of the system deprived of one electron, if accessible (see refs~\citenum{GorGalBae-MP-16,GorBae-arxiv-18} for an in-depth discussion and exceptions).
The kinetic potential is 
\begin{equation}\label{vkin}
v_{\lambda, kin}(\br) = \frac{1}{2}\int \, \, |\nabla_{\br}\Phi_{\lambda}(\sigma,2,...,N|\br)|^2 \md\sigma\md2...\md N,
\end{equation}
and it also  goes usually to zero when $|\br|\to\infty$, as the conditional amplitude in this case becomes insensitive to the position of the reference electron (again, see refs~\citenum{GorGalBae-MP-16,GorBae-arxiv-18} for an in-depth discussion and exceptions).
Finally, the conditional potential is 
\begin{equation}
	\label{vcond}
v_{\lambda, cond}(\br) = \int \sum_{p>1}\frac{1}{r_{1p}} |\Phi_{\lambda}(\sigma,2,...,N|\br)|^2 \md \sigma\md2...\md N,
\end{equation}
and tends manifestly to zero when $|\br|\to\infty$.
For any finite $\lambda$, the difference $E_{0,\lambda}^{N}-E_{0,\lambda*}^{N-1}$ in eq~\ref{ELE} equals minus the exact ionization potential $I_p$ of the physical system, which dictates the asymptotic decay of the density\cite{LevPerSah-PRA-84, AlmBar-PRB-85}
\begin{equation} \label{EE-Ip}
E_{0,\lambda}^{N}-E_{0,\lambda*}^{N-1}=-I_p.
\end{equation}
Similarly, the sum $v_{\lambda, \, eff}(\br)+v^{\lambda}(\br)$ is obviously $\lambda$-independent, as the density is the same for all coupling strengths $\lambda$. It is exactly by equating $v_{\lambda, \, eff}(\br)+v^{\lambda}(\br)$ at $\lambda=0$ and $\lambda=1$ that Baerends and coworkers could derive their insightful decomposition of the KS potential\cite{BuiBaeSni-PRA-89, GriLeeBae-JCP-94, LeeGriBae-ZPD-95, BaeGri-PRA-96, GriLeeBae-JCP-96, BaeGri-JPCA-97}, as this gives an equation for $v^{\lambda=0}$ (i.e. the KS potential) in terms of wavefunction and KS orbitals quantities.


\subsection{General structure of the $\lambda\to\infty$ limit}\label{sec:SCEgeneral}

When $\lambda\to\infty$, the hamiltonian of eq~\ref{AC} has the expansion\cite{SeiGorSav-PRA-07, GorVigSei-JCTC-09, Lew-arxiv-17, CotFriKlu-arxiv-17}
\begin{equation}
	\hat{H}_{\lambda\to\infty}=\lambda(\hat{V}_{ee}+\hat{V}^{SCE})+O(\sqrt{\lambda}),
	\label{eq:haminf}
\end{equation}
where $\hat{V}^{SCE}=\sum_{i=1}^N v^{SCE}(\br_i)$ is the one body potential that makes the classical potential energy operator $\hat{V}_{ee}+\hat{V}^{SCE}$ to have the prescribed ground-state density $\rho(\br)$.\cite{SeiGorSav-PRA-07, GorVigSei-JCTC-09, MalMirCreReiGor-PRB-13}
The modulus squared of the corresponding wavefunction usually collapses into a distribution  that can be written as\cite{Sei-PRA-99, SeiGorSav-PRA-07, GorVigSei-JCTC-09} 
\begin{align}
& | \Psi_{SCE}(1,..., N) | ^2  \nonumber \\
& = \int \frac{\rho(\bs)}{N} \,  \delta(\br_1- \bs) \, \delta(\br_2- \fv_2(\bs))...\delta(\br_N- \fv_N(\bs)) \,  \md\bs,
\label{PsiSCE}
\end{align}
where the co-motion functions $\fv_i(\br)$ describe the perfect correlation between the $N$ electrons. They are non-local functionals of the density satisfying the equation 
\begin{equation}
\rho(\fv_i(\br))\md\fv_i(\br) = \rho(\br)\md \br \qquad (i= 1,\dots, N)
\label{differential}
\end{equation}
which ensures that the probability of finding one electron at position $\br$ in the volume element $\md \br$ be the same of finding electron $i$ at position $\fv_i(\br)$ in the volume element $\md \fv_i(\br)$.
They also satisfy cyclic group properties (for a recent review  on the mathematical properties of the co-motion functions see ref\citenum{SeiDiMGerNenGieGor-arxiv-17}):
\begin{equation}
	\label{eq:groupprop}
\begin{aligned}
\textbf{f}_1(\textbf{r})&\equiv\textbf{r},\\
\textbf{f}_2(\textbf{r})&\equiv\textbf{f}(\textbf{r}),\\
\textbf{f}_3(\textbf{r})&\equiv\textbf{f}\bigl(\textbf{f}(\textbf{r})\bigr),\\
&\ldots\\
\textbf{f}_N(\textbf{r})&=\underbrace{\textbf{f}\bigl(\textbf{f}(\ldots\textbf{f}(\textbf{r})\ldots)\bigr)}_{N-1\text{ times}},\\
&\underbrace{\textbf{f}\bigl(\textbf{f}(\ldots\textbf{f}(\textbf{r})\ldots)\bigr)}_{N\text{ times}}=\textbf{r}.
\end{aligned}
\end{equation}
The corresponding SCE functional, given by\cite{SeiGorSav-PRA-07, MirSeiGor-JCTC-12}
\begin{equation}
	V_{ee}^{SCE}[\rho]=\frac{1}{2}\int \rho(\br)\sum_{i=2}^N\frac{1}{|\br-\fv_i(\br)|}\md\br,
\end{equation}
yields the strong-coupling (or low-density) asymptotic value of the exact Hartree-exchange-correlation functional.\cite{Lew-arxiv-17, CotFriKlu-arxiv-17} Despite the extreme non-locality of $V_{ee}^{SCE}[\rho]$, its functional derivative $v_{Hxc}^{SCE}(\br)=\frac{\delta V_{ee}^{SCE}[\rho]}{\delta \rho(\br)}$ can be computed from the exact force equation\cite{SeiGorSav-PRA-07, MalGor-PRL-12}
\begin{align}
		\nabla v_{Hxc}^{SCE}(\br)=-\sum_{i=2}^{N}\frac{\br-\fv_i(\br)}{\left | \br-\fv_i(\br) \right|^3}.
		\label{eq:pot_sce}
	\end{align}
According to eq~\ref{eq:vextlambda}, the one-body potential $v^{SCE}(\br)$ of eq~\ref{eq:haminf} is exactly equal to minus $v_{Hxc}^{SCE}(\br)$: in fact, the gradient of  $v_{Hxc}^{SCE}(\br)$ represents the net repulsion felt by an electron in $\br$ due to the other $N-1$ electrons at positions $\fv_i(\br)$, while $v^{SCE}(\br)$ appearing in the $\lambda\to\infty$ hamiltonian of eq~\ref{eq:haminf} exactly compensates this net force, in such a way that the classical potential energy operator $\hat{V}_{ee}+\hat{V}^{SCE}$ is stationary (and minimum) on the manyfold parametrized by the co-motion functions. Equation~\ref{eq:pot_sce} defines $v_{Hxc}^{SCE}(\br)$ up to a constant, which is fixed by imposing that both $v_{Hxc}^{SCE}(\br)$ and $v^{SCE}(\br)=-v_{Hxc}^{SCE}(\br)$ go to zero when $|\br|\to\infty$.

The effective equation \ref{ELE} for $\sqrt{\rho(\br)}$ in the SCE limit can be easily understood if we divide both sides by $\lambda\,\sqrt{\rho(\br)}$,
\begin{equation}\label{lambdadivided} 
\begin{aligned}
 -\frac{\nabla^2 \sqrt{\rho(\br)}}{2\lambda \sqrt{\rho(\br)}}+\frac{v_{\lambda,N-1}(\br)}{\lambda} + \frac{v_{\lambda, kin}(\br)}{\lambda}+ \\ v_{\lambda, cond}(\br)+\frac{v^{\lambda}(\br)}{\lambda}  =\frac{1}{\lambda} (E_{0,\lambda}^{N}-E_{0,\lambda*}^{N-1}).
 \end{aligned}
\end{equation}
When $\lambda\to\infty$, we see that the first term in the left-hand-side goes to zero, as the density $\rho(\br)$ does not change with $\lambda$ and it is well behaved, with the exception of the values of $\br$ on top of the nuclear positions ${\bf R}_i$, where the density has a cusp and $\frac{\nabla^2 \sqrt{\rho(\br)}}{\sqrt{\rho(\br)}}$ yields back the Coulombic divergence. Nagy and J\'anosfalvi\cite{NagJan-PM-06} have carefully analyzed the $\lambda\to\infty$ behavior at the nuclear cusps in $\frac{\hat{H}_{\lambda}}{\lambda}$, showing that for all $\lambda$ values the kinetic divergence at a nucleus of charge $Z$ at position ${\bf R}_i$ cancels exactly the external potential $-\frac{Z}{\lambda|\br-{\bf R}_i|}$. We can then safely disregard both the kinetic and the Coulombic divergence in the $\lambda\to\infty$ limit. The other case, which we do not consider here, where this term may diverge is when the KS highest-occupied molecular orbital (HOMO) has a nodal plane that extends to infinity.\cite{GorGalBae-MP-16,AscArmKum-PRB-17,GorBae-arxiv-18}

All the remaining terms, except for $v_{\lambda, kin}(\br)$, will tend to a finite, in general non-zero, limiting value, as they grow linearly with $\lambda$ (for example $v^{\lambda}(\br)\to -\lambda v_{Hxc}^{SCE}(\br)$ of eq~\ref{eq:pot_sce}). Notice that $v_{\lambda, cond}(\br)$ has been already defined with the factor $\lambda$ in front, see eq~\ref{vcond}. The only delicate term is $v_{\lambda, kin}(\br)$ of eq~\ref{vkin}, which contains the gradient of a conditional amplitude that is collapsing into a distribution. Several results in the literature suggest\cite{GorVigSei-JCTC-09, Lew-arxiv-17, GroKooGieSeiCohMorGor-JCTC-17} that this term grows with $\lambda$ only as $\sim\sqrt{\lambda}$, thus still vanishing with respect to the other terms. However, we should keep in mind that no rigorous proof of this statement is available at present. Nonetheless, as shown below, the SCE limit provides a perfectly consistent treatment of the leading order of eq~\ref{ELE} when $\lambda\to\infty$, providing further evidence that the kinetic potential $v_{\lambda, kin}(\br)$ should be subleading in eq~\ref{lambdadivided}. 

\subsection{Conditional probability amplitude and ionization potential at the SCE limit}\label{IpSCE1}
We can now use eq~\ref{PsiSCE} to find the conditional amplitude in the SCE limit and to partition the corresponding effective potential into its two components of eqs~\ref{N-1} and \ref{vcond} (as said, the kinetic part disappears in this limit). Notice that although in eq~\ref{PsiSCE} we have considered only one possible permutation of the $N$ electrons (compare the expression e.g. with eq (14) in ref~\citenum{MirSeiGor-JCTC-12}), this does not affect the derivations below, as explicitly shown in Appendix \ref{app_perm}.  Integrating over $\bs$ we get
\begin{equation}
| \Psi_{SCE}(1,..., N)| ^2= \frac{\rho(\br_1)}{N} \,\delta(\br_2- \fv_2(\br_1))...\delta(\br_N- \fv_N(\br_1)),
\end{equation}
and applying equation \ref{FWF} we find
\begin{equation} \label{CA}
| \Phi_{SCE}(2,...,N|1)| ^2= \delta(\br_2- \fv_2(\br))...\delta(\br_N- \fv_N(\br)).
\end{equation}
Equation \ref{CA} shows that the conditional amplitude gets a very transparent meaning in the SCE limit, as it simply gives the position of the other $N-1$ electrons as a function of the position $\br$ of the first electron. 

In what follows we label with ``SCE'' the terms that survive when we take the limit $\lambda\to \infty$ of  eq~\ref{lambdadivided}. We then use eq~\ref{CA} to evaluate in this limit $v^{SCE}_{N-1}(\br)$,
\begin{equation}\label{almost}
\begin{aligned}
& v^{SCE}_{N-1}(\br)=\\
 & \int \! \left(-\sum^N_{i=2} v_{Hxc}^{SCE}(\br_i) + \! \sum^N_{j>i, \, i=2} \frac{1}{r_{ij}}\right) \!
  \prod_{i=2}^N\delta(\br_i- \fv_i(\br)\!)\md \br_2...\md \br_N \\& - E_{0,SCE*}^{N-1}= \\
 & =- \sum^N_{i=2} v_{Hxc}^{SCE}(\fv_i(\br))  + \sum^N_{j>i=2} \frac{1}{| \fv_i(\br)-\fv_j(\br)|}- E_{0,SCE*}^{N-1} 
\end{aligned}
\end{equation}
Now we use the fact that the ground-state energy of the $N$-particles system with density $\rho(\br)$ at the SCE limit is simply given by the value of the classical potential energy $\hat{V}_{ee}+\hat{V}^{SCE}$ on the manyfold parametrized by the co-motion functions,
\begin{equation}
E_{0,SCE}^{N}= - \sum^N_{i=1}v_{Hxc}^{SCE}(\fv_i(\br)) + \sum^N_{i>j, \, j=1} \frac{1}{| \fv_i(\br)-\fv_j(\br)|},
\end{equation}
which allows us to rewrite the first term of equation \ref{almost} as
\begin{equation}\label{done}
\begin{aligned}
\braket{\Phi_{SCE}(\sigma,2,...,N|\br)|\hat{H}_{SCE}^{N-1} |\Phi_{SCE}(\sigma,2,...,N|\br)} = & \\
 = E_{0,SCE}^{N} + v_{Hxc}^{SCE}(\br) - \sum^N_{i=2} \frac{1}{|\br-\fv_i(\br)|} & \\
\end{aligned}
\end{equation}
The last two terms in the right-hand-side of eq~\ref{done} vanish for $|\br|\to\infty$. On the other hand, by construction  $v_{N-1}(\br)\rightarrow0 $ when $|\br|\to\infty$, and thus necessarily
 \begin{equation}\label{SCEparadox}
 E_{0,SCE}^{N}= E_{0,SCE*}^{N-1},
\end{equation}
and we obtain the final simple expression for $v^{SCE}_{N-1}(\br)$,
\begin{equation} \label{vN-1sce}
v^{SCE}_{N-1}(\br)= v_{Hxc}^{SCE}(\br) - \sum^N_{i=2} \frac{1}{|\br-\fv_i(\br)|}.
\end{equation}
Equation~\ref{SCEparadox} might look puzzling, but one could also expect it from the fact that, as said, in the SCE limit we obtain the quantities that survive in eq~\ref{lambdadivided} when we take the $\lambda\to\infty$ limit. This means that
the difference $E_{0,\lambda}^{N}-E_{0,\lambda*}^{N-1} $ grows linearly with $\lambda$ for large $\lambda$,
\begin{equation}
\lambda\rightarrow\infty \quad  E_{0,\lambda}^{N}-E_{0,\lambda*}^{N-1} \sim \lambda \left( E_{0, SCE}^N - E_{0,SCE*}^{N-1}\right) + O(\sqrt{\lambda}) + ... 
\label{eq:Iplargelambda}
\end{equation}
Then we see that the only way in which eq~\ref{EE-Ip} can be satisfied when $\lambda$ goes to infinity is if eq~\ref{SCEparadox} holds. Indeed this result was already implicit in ref~\citenum{SeiGorSav-PRA-07}, where it was noticed that the configuration with one electron at infinity must belong to the degenerate minimum of the classical potential energy operator $\hat{V}_{ee}+\hat{V}^{SCE}$. Equation~\ref{eq:Iplargelambda} shows that also for the next leading order $\sim\sqrt{\lambda}$ there should be no energy cost to remove one electron, a statement that is implicitly contained in ref~\citenum{GorVigSei-JCTC-09}. 

Notice that the zero ionization energy of eq~\ref{SCEparadox} concerns the $\lambda\to\infty$ hamiltonian in the adiabatic connection of eq~\ref{AC}. A very different result is obtained if $v_{Hxc}^{SCE}(\br)$ is used as an approximation for the Hartree-XC potential in the self-consistent KS equations, where the corresponding KS HOMO eigenvalue has been found to be very close to minus the exact ionization potential for low-density systems,\cite{MalGor-PRL-12,MenMalGor-PRB-14} displaying the correct step structure when the number of electrons is changed in a continuous way.\cite{MirSeiGor-PRL-13}

\section{Different types of response potentials:  $v_{resp}(\br)$, $\overline{v}_{resp}(\br)$, and $v_{resp}^{SCE}(\br)$}
\label{sec:vrespvarious}
In order to compare the SCE response potential with the physical one, we first review the different possible definitions that appear in the literature\cite{BuiBaeSni-PRA-89,  GriLeeBae-JCP-94, LeeGriBae-ZPD-95, GriLeeBae-JCP-96, BaeGri-PRA-96, BaeGri-JPCA-97, VucLevGor-JCP-17} for this term, and fully define the response potential in the SCE limit.

We start from the pair-density $P_{2}^{\lambda}(\br,\br')$, associated to the hamiltonian in eq~\ref{AC} according to the formula:
\begin{equation}\label{Pair}
P_{2}^{\lambda}(\br,\br')=N(N-1) \int |\Psi_{\lambda}(\br\sigma,\br'\sigma',..., N)|^2 \md \sigma\md \sigma' \md \bx_3...\md \bx_N,
\end{equation}
and the corresponding exchange-correlation pair-correlation function $g_{xc}^{\lambda}(\br,\br')$ at a given coupling strength $\lambda$,
\begin{gather}\label{pcf}
g^{\lambda}_{xc}(\br,\br') = \frac{P_{2}^{\lambda}(\br,\br')}{\rho(\br) \rho(\br')}-1
\end{gather}
We also define the coupling-constant averaged (CCA) pair-correlation function $\overline{g}_{xc}(\br,\br')$
\begin{gather}\label{pcfcca}
\overline{g}_{xc}(\br,\br') = \int_0^1 g^{\lambda}_{xc}(\br,\br'), \md\lambda.
\end{gather}
In what follows we use the subscript $s$ when the quantity of interest refers to the KS or $\lambda = 0$ case and we omit the subscript $\lambda$ when it refers to the physical system $\lambda = 1$.

\subsection{Response potential in terms of kinetic and interaction components}\label{Resp1} 
The XC functional of KS DFT can be written as
\begin{equation}
\begin{aligned}
& E_{xc}[\rho] =  T_c[\rho]+ V_{ee}[\rho]-U[\rho]= \\ 
& =\int   v_{c,kin}(\textbf{r})\rho(\textbf{r})\, \md \br + \frac{1}{2} \iint \rho(\textbf{r}) \rho(\textbf{r}') \frac{g_{xc}(\textbf{r} , \textbf{r}')}{| \br - \br' |} \md \br \md \br',
\end{aligned}
\end{equation}
where
\begin{align}\label{vckin}
& v_{c, kin}(\br) =  v_{kin} (\br) - v_{s,kin} (\br) \nonumber= \\ & \!\frac{1}{2}\!\int \! \!\left( |\nabla_\br\Phi(\sigma,2,...,N|\br)|^2\! -\! |\nabla_\br\Phi_s(\sigma,2,...,N|\br)|^2\right) \,\md \sigma\md2...\md N
\end{align}
 If we now take the functional derivative of the XC energy with respect to the density, we can recognise four different contributions to the XC potential,\cite{GriLeeBae-JCP-94, BaeGri-JPCA-97}
 \begin{align}\label{vxcvresp}
&v_{xc}(\textbf{r})=\frac{\delta E_{xc}[\rho]}{\delta\rho(\textbf{r})}\nonumber= \\ &  v_{c,kin}(\textbf{r}) + v_{c,kin}^{resp}(\textbf{r}) +v_{xc,hole}(\textbf{r}) + v_{xc,hole}^{resp}(\textbf{r}),
\end{align}
where 
\begin{equation}\label{vkinresp}
v_{c,kin}^{resp}(\textbf{r})= \int \rho(\textbf{r}')\frac{\delta v_{c,kin}(\textbf{r}')}{\delta\rho(\textbf{r})}\md \br',
\end{equation} 
\begin{equation}
v_{xc,hole}(\textbf{r})= \int\rho(\textbf{r}')\frac{g_{xc}(\textbf{r},\textbf{r}')}{| \br - \br' |} \,\md \br',
\end{equation}
and
\begin{equation}\label{vxcholeresp}
v_{xc,hole}^{resp}(\textbf{r})=\frac{1}{2} \iint  \frac{\rho(\textbf{r}') \rho(\textbf{r}'')}{| \br' - \br'' |} \frac{\delta g_{xc}(\textbf{r}' , \textbf{r}'')}{\delta\rho(\textbf{r})} \md\br' \md\br''.
\end{equation}
We can also group the potentials in eq \ref{vkinresp} and \ref{vxcholeresp} into one total response potential, $v_{resp}(\textbf{r})$,
\begin{equation}
v_{resp}(\textbf{r})= v_{c,kin}^{resp}(\textbf{r}) + v_{xc,hole}^{resp}(\textbf{r})
\label{vresp}
\end{equation}
By inserting the KS Slater determinant and the $\lambda=1$ wavefunction into eq~\ref{N-1} it has been shown\cite{GriLeeBae-JCP-94, BaeGri-JPCA-97} that
\begin{equation}
v_{resp}(\textbf{r})= v_{N-1}(\textbf{r}) - v_{s, N-1}(\textbf{r}).
 \label{respN-1}
\end{equation}

\subsection{Response potential from the coupling-constant averaged XC hole and comparison between $v_{resp}(\br)$ and $\overline{v}_{resp}(\br)$ }\label{Resp2}
The XC energy can be also written in terms of the CCA $\overline{g}_{xc}(\br,\br')$,
\begin{equation}\label{eq:exc}
E_{xc}[\rho]= \frac{1}{2} \iint \rho(\br) \rho(\br')\frac{\overline{g}_{xc}(\br,\br')}{| \br - \br' |} \md\br \md\br',
\end{equation}
as the integration over $\lambda$ allows to recover the kinetic contribution to $E_{xc}[\rho]$.\cite{LanPer-SSC-75, PerLan-PRB-77, GunLun-PRB-76} Taking the functional derivative of eq~\ref{eq:exc} we obtain two terms\cite{GriMenBae-JCP-16}
\begin{align}\label{vxcvrespbar}
\begin{aligned}
v_{xc}(\textbf{r})= & \frac{\delta E_{xc}[\rho]}{\delta\rho(\br)}
= \overline{v}_{xc,hole}(\textbf{r}) + \overline{v}_{resp}(\textbf{r}),
\end{aligned}
\end{align}
where
\begin{equation}
\overline{v}_{xc,hole}(\textbf{r}) = \int \rho(\br') \frac{\overline{g}_{xc}(\br,\br')}{| \br - \br' |} \md\br',
\end{equation}
and
\begin{equation}
\overline{v}_{resp}(\textbf{r})=\frac{1}{2} \iint  \frac{\rho(\br')\rho(\br'')}{| \br - \br' |} \frac{\delta \overline{g}_{xc}(\br', \br'')}{\delta\rho(\textbf{r})} \md\br' \md\br''.
\label{vrespbar}
\end{equation}
Equation~\ref{vrespbar} defines the quantity $\overline{v}_{resp}(\textbf{r})$, but looking at eq~\ref{vxcvrespbar} one can also determine it as:
\begin{equation}
	\label{eq:vbarrespfromdiff}
\overline{v}_{resp}(\textbf{r}) = v_{xc}(\textbf{r}) - \overline{v}_{xc,hole}(\textbf{r}),
\end{equation}
which is how we have computed the response potential in sec~\ref{physres}.
Comparing eqs. \ref{vxcvresp} and \ref{vxcvrespbar}, we have
\begin{align}
&\overline{v}_{xc,hole}(\br) + \overline{v}_{resp}(\br)= \nonumber \\ & v_{c, kin}(\br) + v_{xc, hole}(\br)  + v_{c, kin}^{resp}(\br)+ v_{xc,hole}^{resp}(\br).
\end{align}
intuitively, one would expect that the sum of the response parts of the l.h.s. equals the response part in the r.h.s., and that so do the remainders on both sides. However, this is not true, and in general we have
 \begin{eqnarray}
v_{c, kin}^{resp}(\br)+ v_{xc,hole}^{resp}(\br) & \neq & \overline{v}_{resp}(\br),\label{a1}\\
v_{c, kin}(\br) + v_{xc, hole}(\br) & \neq & \overline{v}_{xc,hole}(\br)\label{a2}.
\end{eqnarray}

\subsection{Response potential for the SCE limit by means of the energy densities}
The two response potentials defined in eq~\ref{vresp} and in eq~\ref{vrespbar} can be both thought of as a measure that answers the question\cite{GriLeeBae-JCP-94, LeeGriBae-ZPD-95, BaeGri-JPCA-97} ``How sensitive is the pair-correlation function on average to local changes in the density?".  Therefore, it seems interesting to ask what happens to it when electrons are perfectly correlated to each other, i.e. in the SCE limit.

From the AC formalism of eq~\ref{AC}, the integrated form of eq~\ref{eq:exc} is
\begin{align}\label{glob-xc-en}
E_{xc}[\rho]= \int _0 ^{1} \mathcal{W}_{\lambda}[\rho] \md\lambda
\end{align}
where $\mathcal{W}_{\lambda}[\rho]$ is the (global) AC integrand, defined as
\begin{equation}\label{eq:w_lam}
\mathcal{W}_\lambda[\rho]= \langle \Psi_{\lambda}  |\hat{V}_{ee}| \Psi_{\lambda} \rangle - U[\rho].
\end{equation}
We can generalise eq~\ref{glob-xc-en} to any XC energy along the adiabatic connection as
\begin{align}
E_{xc}^{\lambda}[\rho]= \int _0 ^{\lambda} \mathcal{W}_{\lambda'}[\rho] \md\lambda'.
\end{align}
Using the expansion of the (global) AC integrand in the strongly-interacting limit\cite{Sei-PRA-99, SeiGorSav-PRA-07, GorVigSei-JCTC-09, Lew-arxiv-17, CotFriKlu-arxiv-17}
\begin{equation}\label{eq:w_ld} \begin{aligned}
\mathcal{W}_\lambda[\rho] &= \mathcal{W}_\infty[\rho] + \mathcal{W}'_\infty[\rho] \lambda^{-1/2} \\
&+ \mathcal{O}(\lambda^{-n}) \quad (\lambda \to \infty \,\, , \,\, n\ge 5/4),
\end{aligned}
\end{equation}
to first order we obtain
\begin{align}\label{1Exc}
\lambda\rightarrow\infty \, \, \, \, \, \, \, \, \, \, \, \,E_{xc}^{\lambda}[\rho]\cong \int _0 ^{\lambda} \mathcal{W}_{\infty}[\rho] \md\lambda'= \lambda \mathcal{W}_{\infty}[\rho]
\end{align}
Defining the SCE XC energy as
\begin{equation} 
	\label{eq:def1ExcSCE}
E_{xc}^{SCE}=\lim_{\lambda\to\infty}\frac{E_{xc}^{\lambda}}{\lambda},
\end{equation}
and inserting eq~\ref{eq:def1ExcSCE} into eq~\ref{1Exc} we get the simple relation
\begin{equation} \label{ExcSCE}
E_{xc}^{SCE}[\rho]=\mathcal{W}_\infty[\rho]=V_{ee}^{SCE}[\rho]-U[\rho]
\end{equation}
In recent years, focus has been brought to the importance of using the local counterpart of the global integrand $\mathcal{W}_{\infty}[\rho]$, i.e. the so-called energy density, $w_\lambda[\rho](\br)$. This different approach is especially important for DFAs (Density Functional Approximations) in view of the fact that
local models are generally more amenable to the construction of size-consistent and accurate methods than their global counterparts.\cite{VucIroWagTeaGor-PCCP-17, VucIroSavTeaGor-JCTC-16, BahZhoErn-JCP-16}
The local analogue of eq~\ref{glob-xc-en} for the XC energy becomes:
\begin{align}
	E_{xc}[\rho]= \int_0^1 \md\lambda \int  \rho(\br)\,w_\lambda[\rho](\br) \, \md \br.
	\label{endens}
\end{align}
Whenever energy densities are used it is crucial to define a ``gauge" within which all the quantities taken into account are computed consistently at different $\lambda$-values, being the choice of $w_\lambda(\br)$ not unique.
A physically sound and commonly used gauge of the energy density is the one given in terms of the electrostatic potential of the XC hole, which corresponds to
\begin{equation}
w_{1}[\rho](\textbf{r}) =  \frac{1}{2}v_{xc,hole}(\br)
\end{equation}
and 
\begin{equation}
\overline{w}[\rho](\textbf{r}) = \frac{1}{2}\overline{v}_{xc,hole}(\br)
\label{eq:wbar}
\end{equation}
where ${w}_1[\rho](\textbf{r})$ in the literature is also labeled as $w[\rho](\textbf{r})$ or $w_{xc}[\rho](\textbf{r})$, while for $\overline{w}[\rho](\textbf{r})$ the symbol  $\overline{w}_{xc}[\rho](\textbf{r})$ or $\epsilon_{xc}[\rho](\br)$ is also commonly used. \\
The corresponding energy density at $\lambda=0$, $w_{0}[\rho](\textbf{r})$, is usually also labeled $\epsilon_{\rm x}(\br)$ or $w_{x}[\rho](\textbf{r})$. For $\lambda \to \infty$ we have, in this gauge, \cite{MirSeiGor-JCTC-12}
	\begin{align}
w_\infty[\rho](\br)=\frac{1}{2}\sum_{i=2}^N\frac{1}{|\br-\fv_i(\br)|}-\frac{1}{2}v_H(\br),
		\label{sce_edens}
	\end{align}
where $v_H(\br)$ is the Hartree potential.
The AC integrand at $\lambda \rightarrow \infty$ can then be written as
\begin{equation} 
\mathcal{W}_{\infty}[\rho]= \frac{1}{2} \iint \rho(\textbf{r}) \rho(\textbf{r}') \frac{g_{xc}^{\infty}(\textbf{r} , \textbf{r}')}{| \br - \br' |} \md \br \md \br'
\end{equation}
Taking the functional derivative of $\mathcal{W}_{\infty}[\rho]$ w.r.t. the density we obtain
\begin{equation}
v_{xc}^{SCE}(\textbf{r})= \frac{\delta \mathcal{W}_{\infty}[\rho]}{\delta\rho(\textbf{r})}= v_{xc,hole}^{SCE}(\textbf{r}) + v^{SCE}_{resp}(\textbf{r}),
\end{equation}
where
\begin{equation}
  v_{xc,hole}^{SCE}(\textbf{r})=\int \rho(\textbf{r}')\frac{g_{xc}^{\infty}(\textbf{r} , \textbf{r}')}{| \br - \br' |} \md \br'=2 \, w_{\infty}(\textbf{r}),
\end{equation}
and
\begin{equation}
v^{SCE}_{resp}(\textbf{r})= \frac{1}{2} \iint  \frac{\rho(\textbf{r}') \rho(\textbf{r}'')}{| \br' - \br'' |} \frac{\delta g_{xc}^{\infty}(\textbf{r}' , \textbf{r}'')}{\delta\rho(\textbf{r})} \md\br' \md\br''.
\end{equation} 
Finally, inserting the explicit expression for the energy density for $\lambda \to \infty$ (eq~\ref{sce_edens}), we find the response potential at the SCE limit
\begin{align}\label{vrespsce}
v_{resp}^{SCE}(\br) =   v_{xc}^{SCE}(\textbf{r}) -2 \, w_{\infty}(\textbf{r}) =  v^{SCE}_{Hxc}(\textbf{r}) -  \sum_{k=2}^N
\frac{1}{|\textbf{r} - \textbf{f}_k(\textbf{r})|},
\end{align}
which is exactly equal to $v_{N-1}^{SCE}(\br)$ of eq~\ref{vN-1sce}.
Notice that the SCE response potential of eq~\ref{vrespsce} scales linearly with respect to uniform scaling of the density:\cite{LevPer-PRA-85}
	\begin{equation}\label{scalingbeh}
	v_{resp}^{SCE}(\br)[\rho_{\gamma}] = \gamma\, v_{resp}^{SCE}(\gamma\,\br)[\rho], 
		\end{equation}
where $\rho_{\gamma}(\br) \equiv \gamma^3 \rho(\gamma\,\br)$ is a scaled density.  

\subsubsection{SCE response potential for a two-electron density}\label{2el-case}
When the number of electrons equals two, we also have another expression for computing $v_{resp}^{SCE}(\br)$.
In this case the SCE total energy $E^{N=2}_{0,SCE}$ of sec~\ref{IpSCE1} is equal to
\begin{equation}
 E_{0,SCE}^{N=2}=\frac{1}{|\br - \fv(\br)|} - v_{Hxc}^{SCE}(\br) -v_{Hxc}^{SCE}(\fv(\br)) .
\end{equation}
where the r.h.s. is the value of the SCE potential energy on the manifold parametrized by the co-motion function. This value is a degenerate minimum, meaning that we can evaluate it at any point lying on the manifold, such as for $|\textbf{r}| \rightarrow \infty $ (for a nice illustration of the degenerate minimum of the SCE potential energy, the interested reader is addressed to fig 1 of ref~\citenum{GroKooGieSeiCohMorGor-JCTC-17}). 
When $|\br|\to\infty$, the potential  $v_{Hxc}^{SCE}(\br)$ is gauged to go to zero. At the same time, the co-motion function $\fv(\br)$ will tend to a well defined position $\br_0$ well inside the density, i.e. $\fv(\br\to\infty) \to \br_0$. We thus have
\begin{equation}\label{vSCE0}
\frac{1}{|\br - \fv(\br)|} - v_{Hxc}^{SCE}(\br) -v_{Hxc}^{SCE}( \fv(\br)) = - v_{Hxc}^{SCE}(\br_0).
\end{equation} 
Combining eqs~\ref{vrespsce} and \ref{vSCE0} we find
\begin{equation}
 v_{resp}^{SCE} (\br)= - v_{Hxc}^{SCE}( \fv(\br)) + v_{Hxc}^{SCE} (\br_0).
\label{vrespsce-2}
 \end{equation}

\section{Examples of CCA and SCE response potentials}\label{physical}
We have computed the SCE response potential, $v_{resp}^{SCE}(\br)$, for small atoms and for the hydrogen molecule at equilibrium distance; for this latter case and for the species H$^-$, He, Be, and Ne also accurate CCA response potentials $\overline{v}_{resp}(\br)$ have been obtained. Notice that, in previous works, several authors\cite{BuiBaeSni-PRA-89, GriLeeBae-JCP-94, LeeGriBae-ZPD-95, BaeGri-PRA-96, GriLeeBae-JCP-96, BaeGri-JPCA-97, RyaSta-JCP-14, CueAyeSta-JCP-15, CueSta-MP-16, KohPolSta-PCCP-16, RyaOspSta-JCP-17} have computed the response potential at physical coupling strength, $v_{resp}(\br)$ of eqs~\ref{vresp}-\ref{respN-1}. To our knowledge, accurate CCA response potentials $\overline{v}_{resp}(\br)$ (eq~\ref{vrespbar}) are reported here for the first time.
In Appendix~\ref{app_datavalidat} we also briefly discuss the extent of the error resulting from combining data coming from different methods, namely from the Lieb Maximisation procedure \cite{TeaCorHel-JCP-09, TeaCorHel-JCP-10, VucIroSavTeaGor-JCTC-16} and Hylleraas-type wavefunctions \cite{MirUmrMorGor-JCP-14, FreHuxMor-PRA-84} or Quantum Monte Carlo calculations \cite{UmrGon-PRA-94, FilGonUmr-INC-96, MirUmrMorGor-JCP-14} as explained in the next sections.
In all the figures all quantities are reported in atomic units.
\subsection{Computational details for the atomic densities}
For the sake of clarity, we treat in separate sections the computation of $v_{resp}^{SCE}(\br)$ and $\overline{v}_{resp}(\br)$ for atoms. 
\subsubsection{SCE response potential}
 The calculation of $v_{resp}^{SCE}(\br)$ for spherical atoms is based on the ansatz for the radial part of the co-motion functions reported in ref~\citenum{SeiGorSav-PRA-07}. These co-motion functions are exact for $N=2$,\cite{ButDepGor-PRA-12} and for $N>2$ they give either the exact SCE solution or get very close to it.\cite{SeiDiMGerNenGieGor-arxiv-17} Moreover, even when they are not truly optimal, the corresponding potential still satisfies eq~\ref{eq:pot_sce}.\cite{SeiDiMGerNenGieGor-arxiv-17}  This means that we are  in any case using a perfectly correlated wavefunction to compute a meaningful response potential. The radial co-motion functions $f_i(r)$ of ref~\citenum{SeiGorSav-PRA-07} are given
\begin{equation}\label{comff}
\begin{aligned}
\text{for odd \(N\)}, & \quad k=1 \cdots  \frac{N-1}{2}\\
f_{2k+1}(r)= &\begin{cases} N_{e}^{-1}[2k+N_{e}(r)] \qquad \qquad r\leq{a_{N-2k}} \\ N_{e}^{-1}[2N-2k-N_{e}(r)] \qquad r>a_{N-2k}\end{cases};
\\
\text{for even \(N\)}, & \quad k=1 \cdots  \frac{N-2}{2}\\
f_{2k}(r)=& \begin{cases} N_{e}^{-1}[2k-N_{e}(r)] \qquad r\leq{a_{2k}} \\ N_{e}^{-1}[N_{e}(r)-2k] \qquad r>a_{2k}\end{cases}\\
& f_N(r)=N_{e}^{-1}[N-N_{e}(r)] \\
\end{aligned}
\end{equation}
where $N$ is the number of electrons, $N_e(r)$  is the cumulant function, 
\begin{equation}\label{cumulant}
 N_e(r)= \int _0^r 4\pi x^2\, \rho(x)\, \md x ,
\end{equation} $N_e^{-1}(y)$ its inverse, defined for $y \in [0, N) $, and $a_i$ are the (radial) distances for which $N_e (a_i) = i$, with $i$ integer.
These radial co-motion functions give the distances from the nucleus of the remaining $N-1$ electrons as a function of the distance $r$ of the first one. The relative angles between the electrons are found by minimizing the total repulsion energy for each given $r$.\cite{SeiGorSav-PRA-07, MenMalGor-PRB-14}
The SCE potential, $v_{Hxc}^{SCE}(r)$,  is then obtained by integration of eq~\ref{eq:pot_sce}. Finally, we apply eq~\ref{vrespsce} (or, equivalently for $N=2$, eq~\ref{vrespsce-2}) to get the SCE response potential.

This procedure is very `robust' meaning that we have obtained comparable SCE response potentials using densities of different levels of accuracy. The densities we have used were obtained from
\begin{enumerate} [label=(\Alph*)]
\item  CCSD calculations and aug-cc-pCVTZ basis set stored on a 0.01 bohr grid, see ref~\citenum{VucIroSavTeaGor-JCTC-16},
\item  Hylleraas-type wave functions, see refs~\citenum{FreHuxMor-PRA-84, MirUmrMorGor-JCP-14}, for the two-electron systems and Quantum Monte Carlo calculations, see refs~\citenum{UmrGon-PRA-94, FilGonUmr-INC-96, AlsResUmr-PRA-98}, for the others.
\end{enumerate}
The cumulant function of eq~\ref{cumulant} was computed either with simple interpolations between the gridpoints of a given density or in some cases (for H$^-$, He, and Li$^+$) with explicitly fitted densities, constrained to satisfy the cusp condition and the correct asymptotic behaviour. \\
Group (A) regards all the systems taken into account. Group (B) regards the species: H$^-$, He, Be, and Ne. The figures in sec~\ref{physres} only show the SCE response potential coming each time from the most accurate available density.

\subsubsection{Coupling-constant averaged response potential}
 The equation used in practice to compute $\overline{v}_{resp}(\br)$  is 
\begin{equation}\label{vrespbarprac}
\overline{v}_{resp}(\br)= v_{xc}(\br) - 2\,\overline{w}(\br)
\end{equation}
where $\overline{w}(\br)$ is given in eq~\eqref{eq:wbar}, and was calculated by averaging the energy densities $w_{\lambda}(\br)$ obtained through the Lieb Maximisation procedure and taken from refs~\citenum{IroTea-MP-15,VucIroSavTeaGor-JCTC-16, VucIroWagTeaGor-PCCP-17}, over the interval $[0, 1]$  with an increment $\Delta\lambda = 0.1$ at each $\br$. The XC potentials were taken instead from Hylleraas-type calculations\cite{MirUmrMorGor-JCP-14} or Quantum Monte Carlo results, \cite{UmrGon-PRA-94, FilGonUmr-INC-96, MirUmrMorGor-JCP-14} as they were overall more accurate. This choice is further validated in Appendix~\ref{app_datavalidat}.

\subsection{Computational details for the hydrogen molecule}
 
 For the hydrogen molecule a different approach -- i.e. the ``dual Kantorovich formulation" in the framework of optimal transport theory \cite{ButDepGor-PRA-12, VucWagMirGor-JCTC-15} -- was used for the computation of the SCE potential and thus of the SCE response potential. The basic idea relies on finding the SCE potential as a result of a nested optimization on a parametrized expression which has the correct asymptotic behaviour, the correct cylindrical symmetry and models the barrier region in the midbond. From the optimized potential one derives the co-motion function by inverting eq~\ref{eq:pot_sce}; for details see ref~\citenum{VucWagMirGor-JCTC-15}. \\
For the CCA energy density, $\overline{w}(\br)$, exactly the same procedure described for atoms has been used. \\
The XC potential for the physical system in this case was obtained within the Lieb Maximisation procedure itself as in ref~\citenum{VucIroSavTeaGor-JCTC-16}, namely as the optimized effective potential that keeps the density fixed minus the Hartree potential and the potential due to the field of the nuclei (see also Appendix~\ref{app_datavalidat} for data validation).

\subsection{Results and discussion}\label{physres}
\begin{figure}
\includegraphics[scale=0.45]{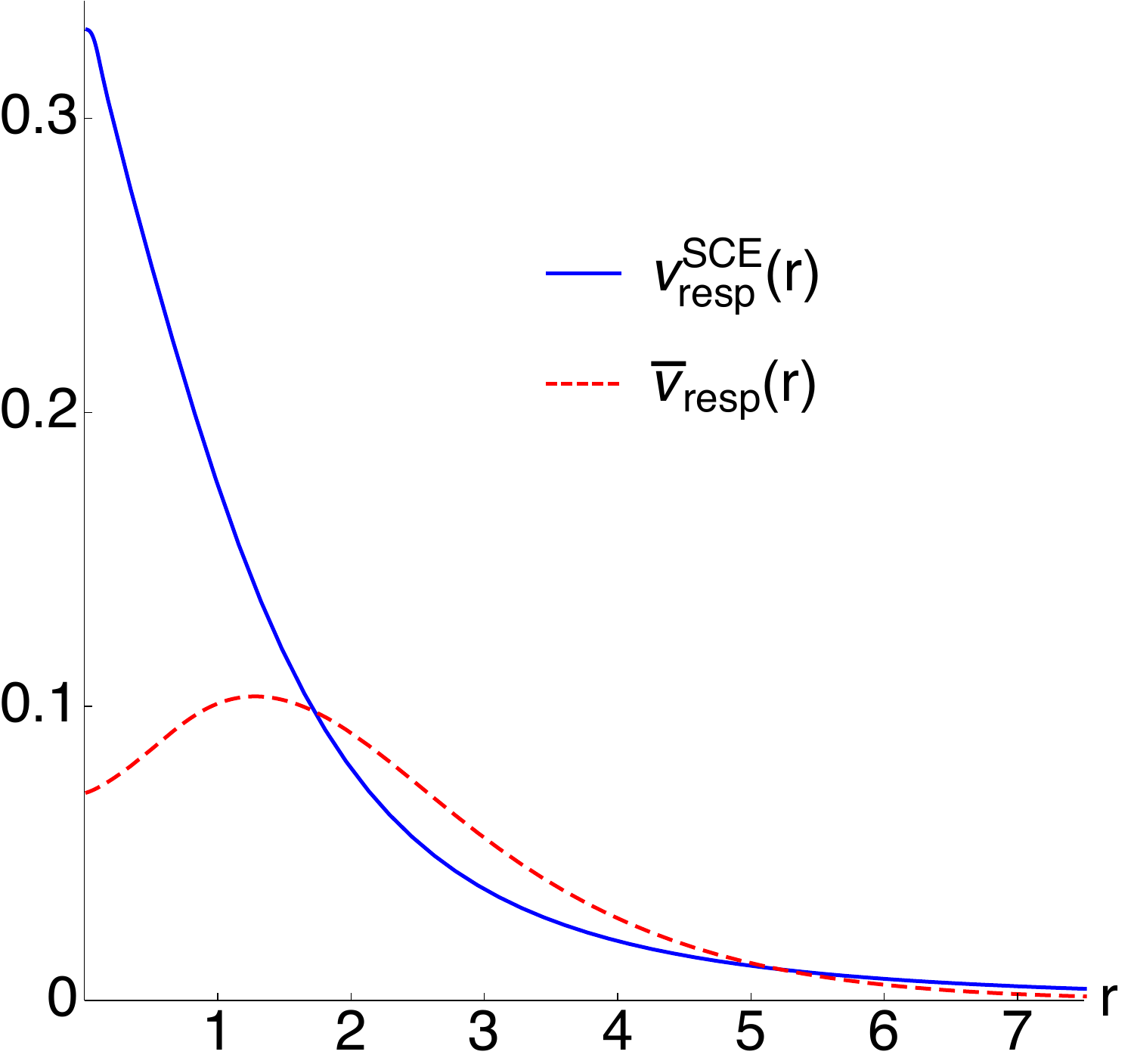}
\caption{Comparison between $\overline{v}_{resp}(r)$ and $v_{resp}^{SCE}(r)$ for the H$^-$ anion. }
\label{Hm-vis-comp}
\end{figure}

We start by showing in fig~\ref{Hm-vis-comp} the  CCA and SCE response potentials for the H$^-$ anion: we see that on average the SCE response potential is larger than the CCA one, but there is an intermediate region, in the range $1.7 \lesssim r \lesssim 5.2 $, where the CCA values are above the SCE ones. Since the SCE response potential does not contain any information on how the kinetic potential is affected by a change in the density, this could be a region where the contribution coming from the kinetic correlation response effects overcome the Coulomb correlation ones, even though we cannot exclude that already the mere Coulombic contribution to correlation is higher in the physical case. Indeed, it has been shown that the SCE pair density can be insensitive to changes in certain regions of the density.\cite{LanMarGerLeeGor-PCCP-16}

\begin{figure}
\includegraphics[scale=0.45]{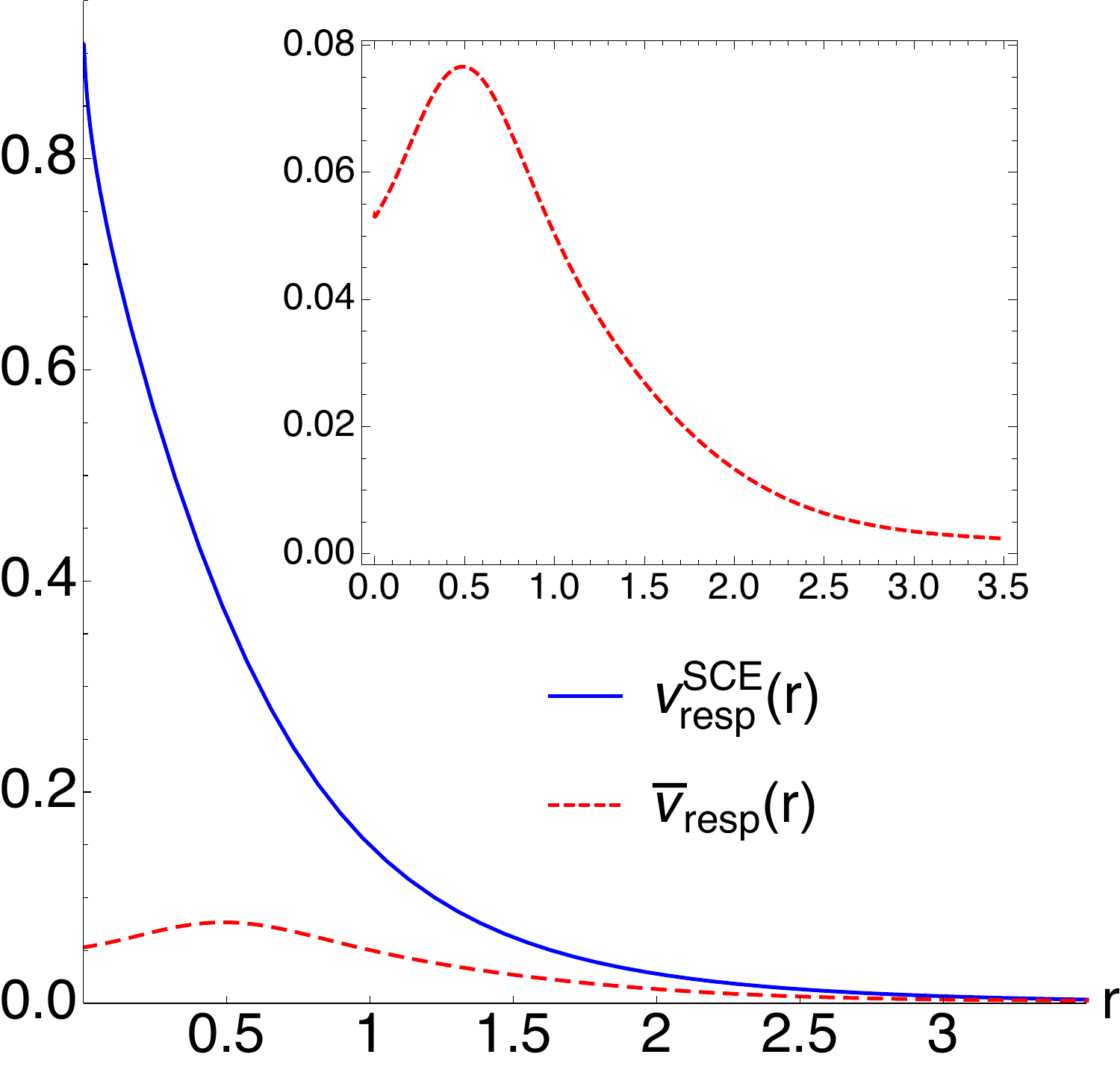}
\caption{Comparison between $\overline{v}_{resp}(r)$ and $v_{resp}^{SCE}(r)$ for the He atom. In the top-right insertion the CCA response potential of He is zoomed in to allow a closer comparison with its response potential at full coupling strength, $v_{resp}(r)$, shown in fig. 3(c) of ref~\citenum{BaeGri-JPCA-97}.}
\label{He-visualcomparison}
\end{figure}

In fig~\ref{He-visualcomparison} we report a similar comparison for the He atom density. Since He is less correlated than H$^-$, in this case the CCA potential $\overline{v}_{resp}(r)$ differs even more from the SCE one. Comparing the two species H$^-$ and He among each other, one can further observe that the value of the distance at which the response potential of the species $i$ has a maximum, $r_{M }^i$, is also shifted leftward (closer to the nucleus) when going from $Z=1$ to $Z=2$, reflecting the contraction of the density. This information is also mirrored in the SCE limit by the shift in the $a_1$ values appearing in equation \ref{comff} for the computation of the co-motion functions for the two species. Indeed we find that $\frac{a_1^{H-}}{a_1^{He}}=\frac{r_{M}^{H-}}{r_{M}^{He}}$.

\begin{figure} [h]
\includegraphics[scale=0.4]{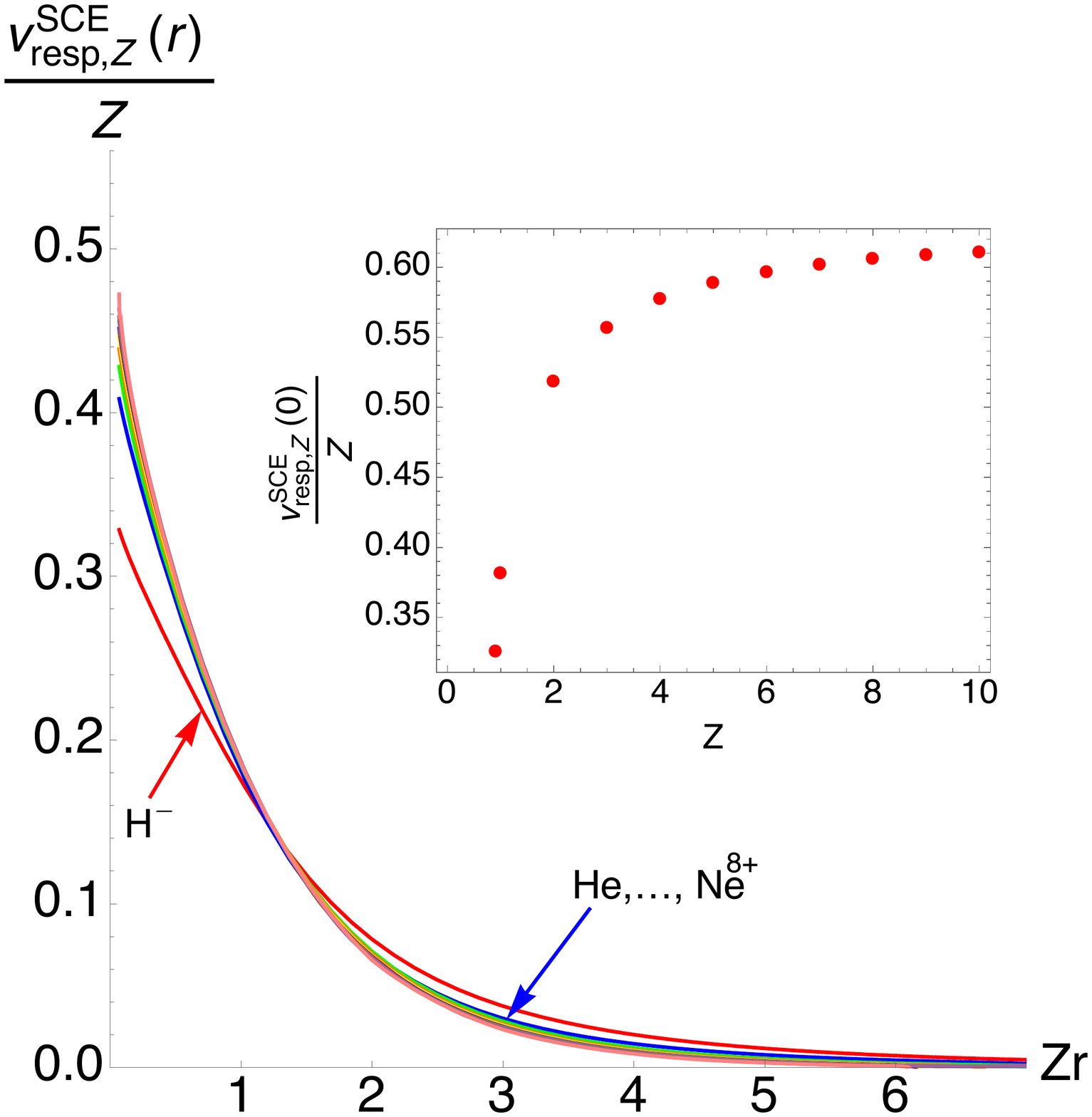}
\caption{Scaled SCE response potentials, $\frac{v_{resp,Z}^{SCE}(\br)}{Z}$ as a function of the scaled coordinate $Z\, r$ for the He series from H$^-$ up to Ne$^{8+}$. 
In the inset, in which only the ``slice" at $r=0$ (i.e. the maximum values of the SCE response potentials) is plotted as a function of the nuclear charge $Z$, also the hypothetical system with $Z= Z_{crit}$ (see text) is considered. 
}
\label{vreSCE-2el-scaled}
\end{figure}

As it could be expected from eq~\ref{scalingbeh}, the response potential at the SCE limit shows an almost perfect scaling behaviour along the He series when we increase the nuclear charge $Z$. This is shown in fig~\ref{vreSCE-2el-scaled}, where we report the scaled potentials, $\frac{v_{resp, Z}^{SCE}(r)}{Z}$ as a function of the scaled coordinate $Z\, r$.  More diffuse densities, like He and H$^-$, deviate from the linear-scaling trend, showing increasing correlation effects in their densities. Such correlation effects (curve lying below the uniformly scaled trend for small  $r$  and above for large $r$ ) are stronger closer to the nucleus.
In the top-right inset of this figure, we show only the values of the maxima of the SCE response potential of each species divided by its nuclear charge, $\frac{v_{resp, Z}^{SCE}(0)}{Z}$ as a function of $Z$. In this inset also a hypothetical system with nuclear charge $Z_{crit}= 0.9110289 $, the minimum nuclear charge that can still bind two electrons (see ref~\citenum{MirUmrMorGor-JCP-14}), is included.

\begin{figure}
\includegraphics[scale=0.3]{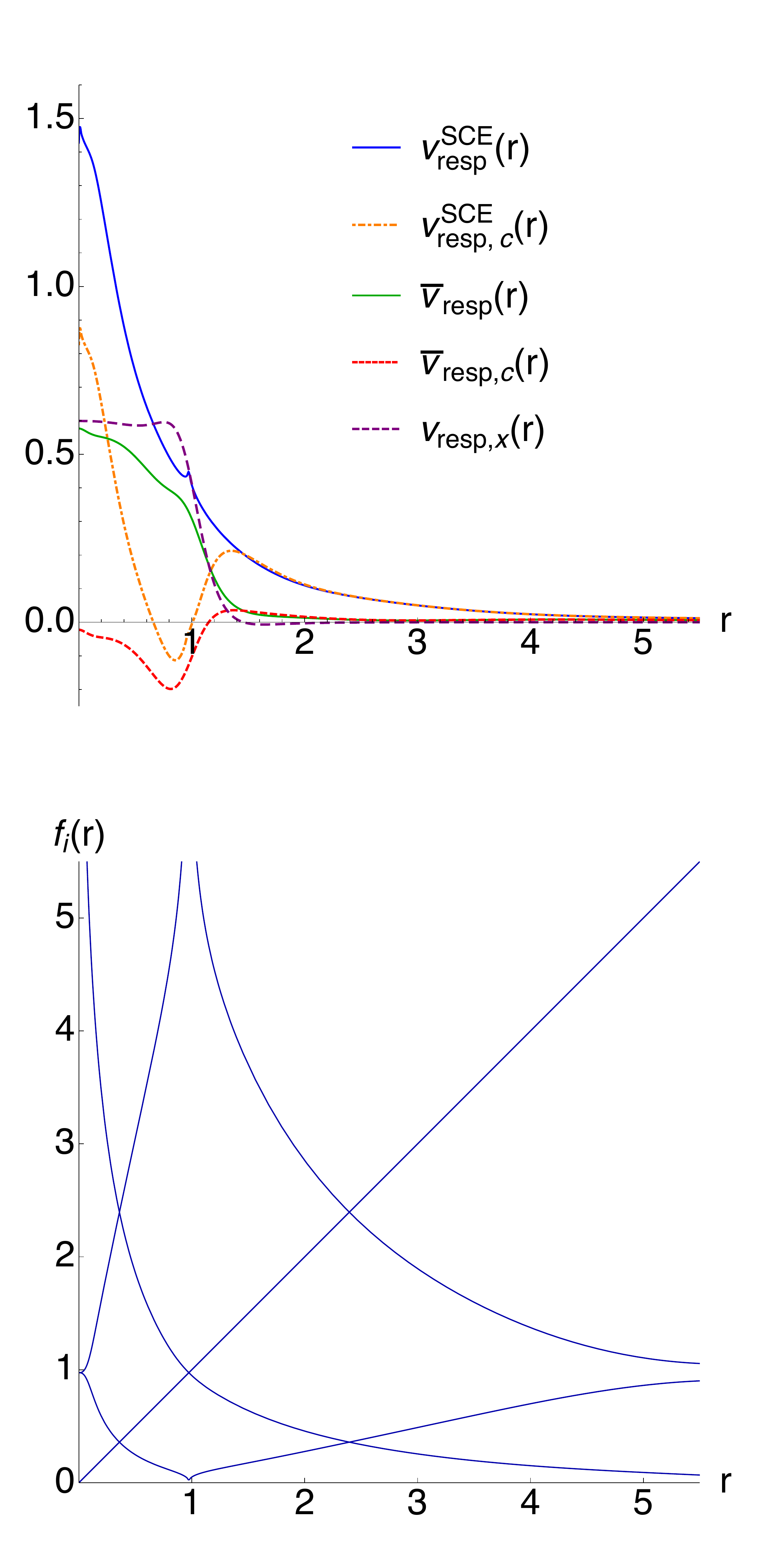}
\caption{Total response potentials $\overline{v}_{resp}(r)$ and $v_{resp}^{SCE}(r)$, and their components $v_{resp,x}(r)$, $\overline{v}_{resp,c}(r)$ and $v_{resp,c}^{SCE}(r)$ (upper panel) and radial co-motion functions (lower panel) for the Be atom.}
\label{Be-vrespSCE-comff}
\end{figure}

In the upper panel of fig~\ref{Be-vrespSCE-comff}  we show the SCE and the CCA response potentials for the Be atom together with the exchange contribution $v_{resp,x}(r)$ (corresponding to $\lambda=0$), and the correlation contributions obtained by subtracting $v_{resp,x}(r)$ from $\overline{v}_{resp}(r)$ and $v_{resp}^{SCE}(r)$. As it was found in ref~\citenum{LeeGriBae-ZPD-95}, the exchange-only response potential shows a clear step structure in the region of the shell boundary. 
The total CCA response potential also shows a step at the same position, while the SCE response potential has a kink.
The kink can be understood by looking at the shape of the radial co-motion functions (see eq~\ref{comff} and the lower panel of the same figure), which determine the structure of the SCE response potential according to eq~\ref{vrespsce}.
The SCE reference system correlates two adjacent electron positions in such a way that the density between them exactly integrates to 1, therefore the $a_i$ appearing in equation \ref{comff} are simply the shells that contain always one electron each.\cite{MalMirGieWagGor-PCCP-14}
For the case of Be, the kink appears at the corresponding  $a_2$-value, which is very close to the shell boundary. In fact, when the reference electron is at distance $r\approx a_2$ from the nucleus, a second electron is found at this same distance (but on the opposite side with respect to the nucleus), while the third electron is very close to the nucleus and the fourth is almost at infinity. This situation results in an abrupt change of the pair density for small variations of the density, as particularly the position of the fourth electron changes very rapidly with small density variations.
Another interesting feature we can observe from fig~\ref{Be-vrespSCE-comff} is that the Coulomb correlation contribution to the CCA response potential, $\overline{v}_{resp,c}(r)$, appears to be negative inside the entire 1s shell region.
Furthermore, while the total physical response potential is always below the SCE one, the exchange part appears to be higher in a region quite close to the shell boundary ($0.6 \lesssim  r  \lesssim 1.0 $). This results in the Coulomb correlation contribution for the SCE-limit case, $v_{resp,c}^{SCE}(r)$, to be also negative in that region.

\begin{figure}
\includegraphics[scale=0.3]{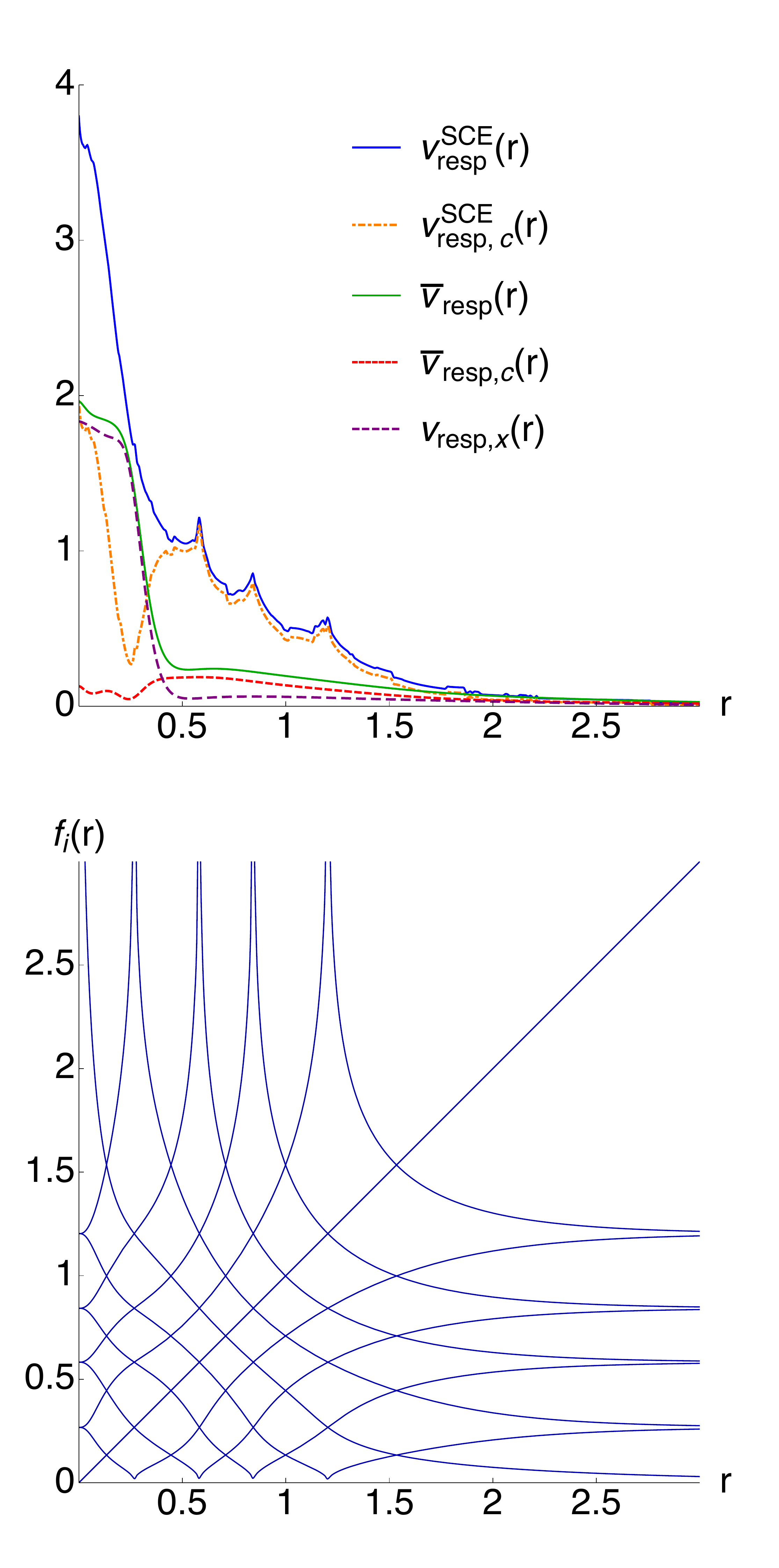}
\caption{Total response potentials $\overline{v}_{resp}(r)$ and $v_{resp}^{SCE}(r)$, and their components $v_{resp,x}(r)$, $\overline{v}_{resp,c}(r)$ and $v_{resp,c}^{SCE}(r)$ (upper panel) and radial-co-motion functions (lower panel) for the Ne atom.}
\label{Ne-vrespSCE-comff}
\end{figure}

\begin{figure}[htbp]
\includegraphics[scale=0.4]{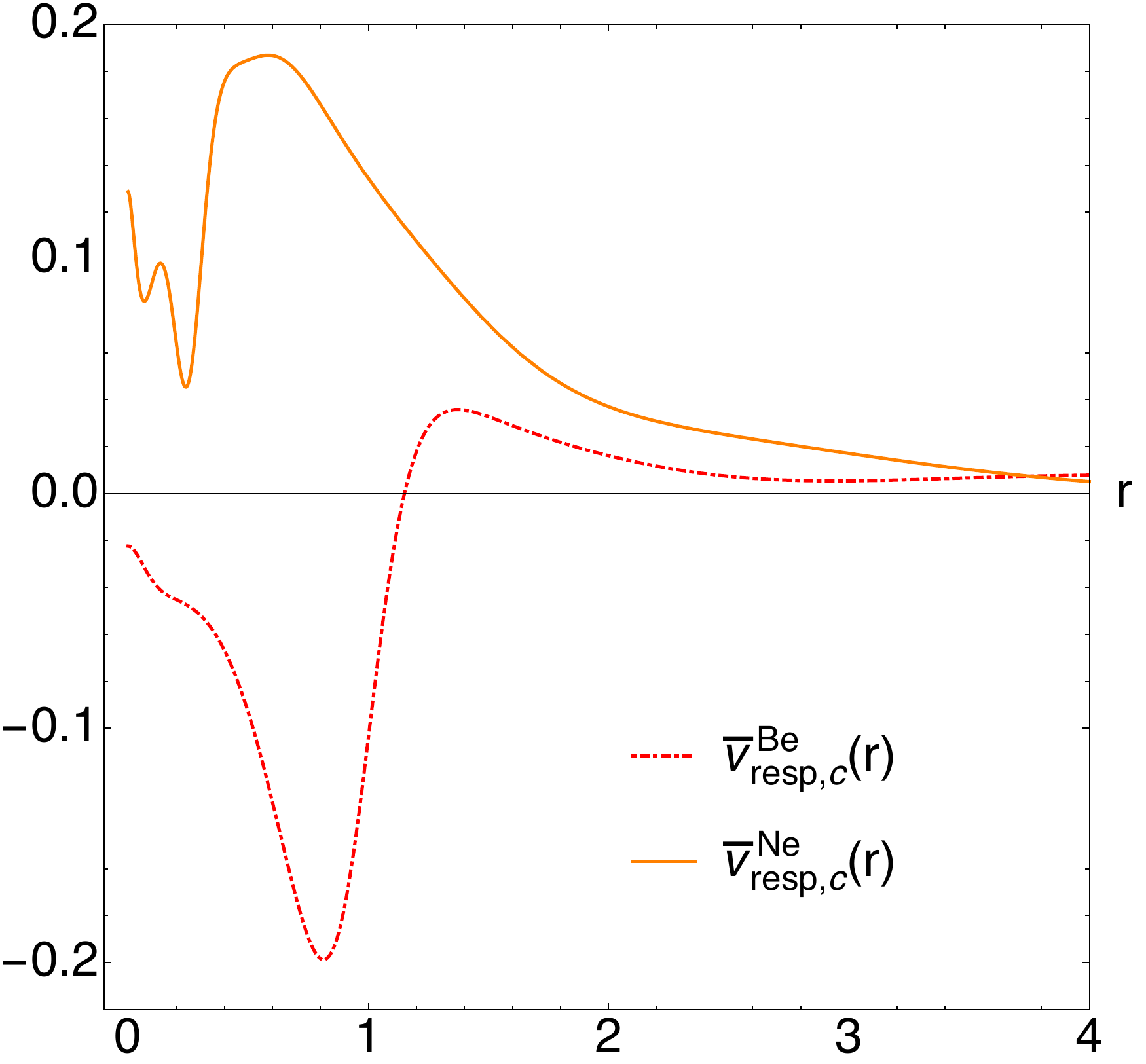}
\caption{Correlation parts of the CCA response potential, $\overline{v}_{resp,c}$, for the Be and the Ne atoms.}
\label{comparisonVreCNeBe}
\end{figure}

In the upper panel of fig~\ref{Ne-vrespSCE-comff} we show the SCE response potential and its correlation part for the Ne atom. The SCE response potentials $v_{resp}^{SCE} (r)$ and $v_{resp, c}^{SCE} (r)$ are numerically less accurate, due to the higher dimensional angular minimization. Nevertheless, the relation between its structure and the corresponding co-motion functions in the lower panel of fig~\ref{Ne-vrespSCE-comff} is clearly visible. We also show the CCA response potentials together with the separate exchange and correlation contributions. Differently from the Be atom, neither the total response potential nor any single correlation contribution (CCA or SCE) is anywhere negative. Still the structure is very similar, showing two steps in the  $v_{resp,x}(r)$ one very tiny at around 0.1  and another at around 0.4 distance from the nucleus and two wells in the $\overline{v}_{resp,c}(r)$. In fig~\ref{comparisonVreCNeBe} we show only the CCA correlation contributions to the CCA response potential of the two species for closer comparison.

\begin{figure}[htbp]
\includegraphics[scale=0.45]{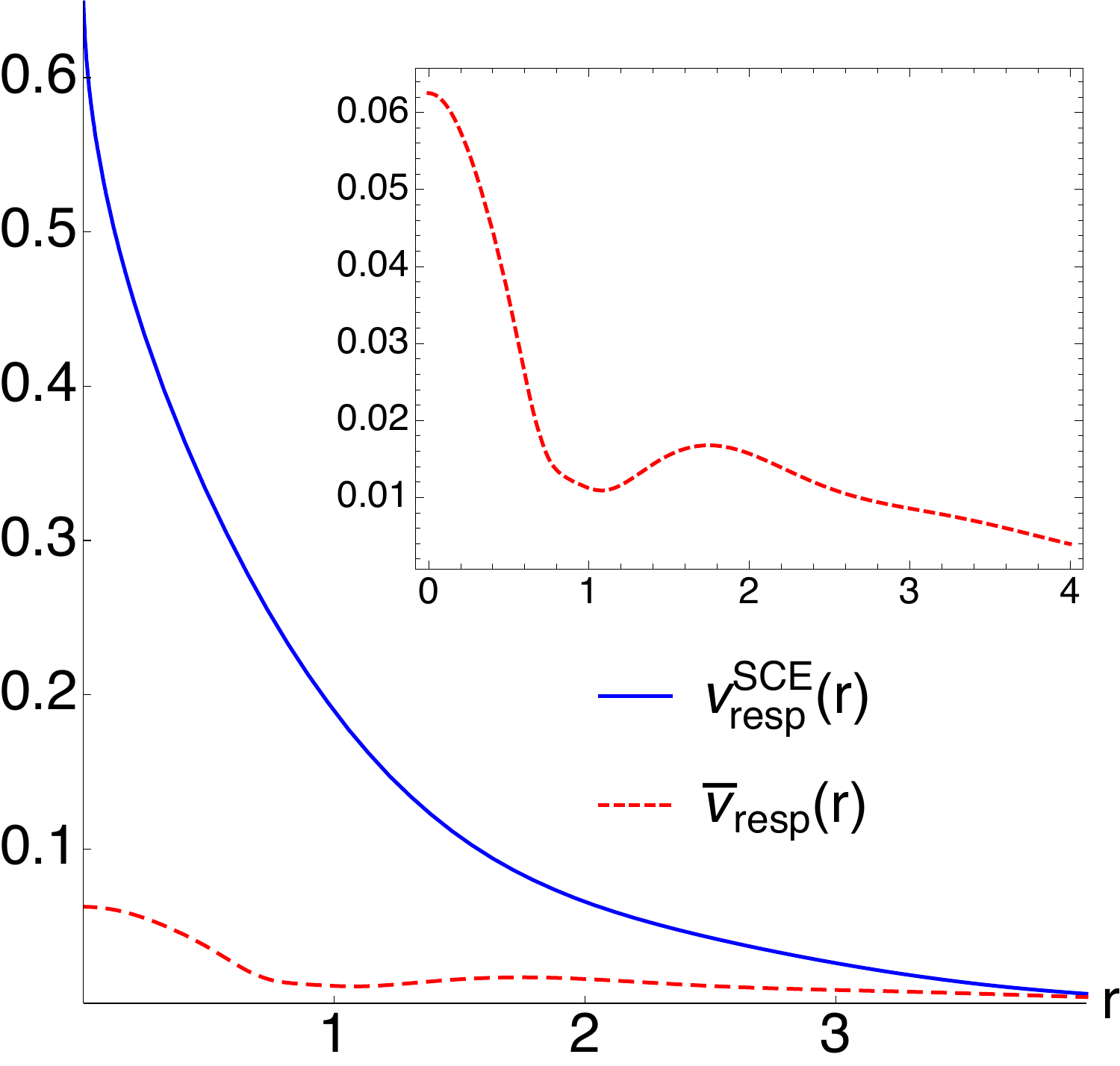}
\caption{Comparison between $\overline{v}_{resp}(\br)$ and $v_{resp}^{SCE}(\br)$ for the H$_2$ molecule at the equilibrium distance along the internuclear axis, origin of the axes being at the bond midpoint. In the top-right insertion the CCA response potential of H$_2$ is zoomed in to allow a closer comparison with its response potential, $v_{resp}(\br)$, shown in fig. 3(a) of ref~\citenum{BaeGri-JPCA-97}.}
\label{vrespH2}
\end{figure}

In fig~\ref{vrespH2} the CCA response potential for the hydrogen molecule at equilibrium distance is shown, together with the SCE one. It is interesting to compare this figure with fig~3(a) of ref~\citenum{BaeGri-JPCA-97}, where the response potential $v_{resp}(\br)$ of eq~\ref{vresp} was reported, together with other components of the XC potential. The response potential at full coupling strength for the  same system is also shown in fig~4 of ref~\citenum{RyaSta-JCP-14}, albeit a minus sign and a constant shift. The overall structure is completely different: in the case shown here there is a local minimum of $\overline{v}_{resp}(\br)$ at approximately 1 bohr distance from the bond midpoint, while $v_{resp}(\br)$  shown in refs~\citenum{BaeGri-JPCA-97, RyaSta-JCP-14} has a maximum located at the nuclei. This must necessarily be due to the coupling-constant average procedure, in which the response of the kinetic and coulombic contribution are taken into account in two different ways. It is then important to keep these different features in mind when one wants to model the response potential,  depending on whether the target is $\overline{v}_{resp}(\br)$ or $v_{resp}(\br)$. 

\section{Simple model for a stretched heteronuclear dimer}\label{sec:ABmodel}
The purpose of this section is to analyse the response potential in the SCE limit for the very relevant case of a dissociating heteroatomic molecule, where the exact response potential is known to develop a characteristic step structure.\cite{AlmBar-INC-85, BaeGri-PRA-96, BaeGri-JPCA-97, TemMarMai-JCTC-09, HelTokRub-JCP-09, BenPro-PRA-16, HodRamGod-PRB-16} Although numerically stable KS potentials have been presented and discussed in the literature for small molecules,\cite{CueAyeSta-JCP-15, HodKraSchGro-JPCL-17} an accurate calculation of the SCE potential for a stretched heterodimer is still not available. In fact, while with the dual Kantorovich procedure\cite{MenLin-PRB-13, VucWagMirGor-JCTC-15} it is possible to obtain accurate values of $V_{ee}^{SCE}[\rho]$ for small molecules, the quality of the corresponding SCE potentials, particularly in regions of space where the density is very small, is not good enough to allow for any reliable analysis.

We then used a simplified one-dimensional (1D) model system, where only the two valence electrons involved in the stretched bond are treated explicitly. Several authors have used this kind of 1D models, which have been proven to reproduce and allow to understand the most relevant features appearing in the exact KS potential of real molecules.\cite{TemMarMai-JCTC-09, HelTokRub-JCP-09, BenPro-PRA-16, HodRamGod-PRB-16}
Here we approximate the density of the very stretched molecule as just the sum of the two ``atomic'' densities
\begin{align}\label{rhomod}
\rho(x) = \rho_a\left(x-\frac{R}{2}\right) + \rho_b\left(x + \frac{R}{2}\right) = \nonumber \\ \frac{a}{2} e^{-a |x - \frac{R}{2}|} + \frac{b}{2} e^{-b |x + \frac{R}{2}|},
\end{align}
where $a$ and $b$ mimic the different ionization potentials of the ``atoms'' (pseudopotentials or frozen cores) and the density is normalized to 2. We have chosen $a > b$, therefore the most electronegative atom will be found to the right side of the origin (at a distance $+\frac{R}{2}$ from it) and the least electronegative to the left.\\
In the last part of this section (subsection \ref{careful}), we inspect and reveal further features of the response potential also at physical coupling strength and put them closely in relation with the SCE scenario discussed in the first part; this investigation is indeed made possible thanks to the simplicity of the model.

\subsection{SCE response potential for the model stretched heterodimer}
\begin{figure} 
\centering
 \includegraphics[scale=0.3]{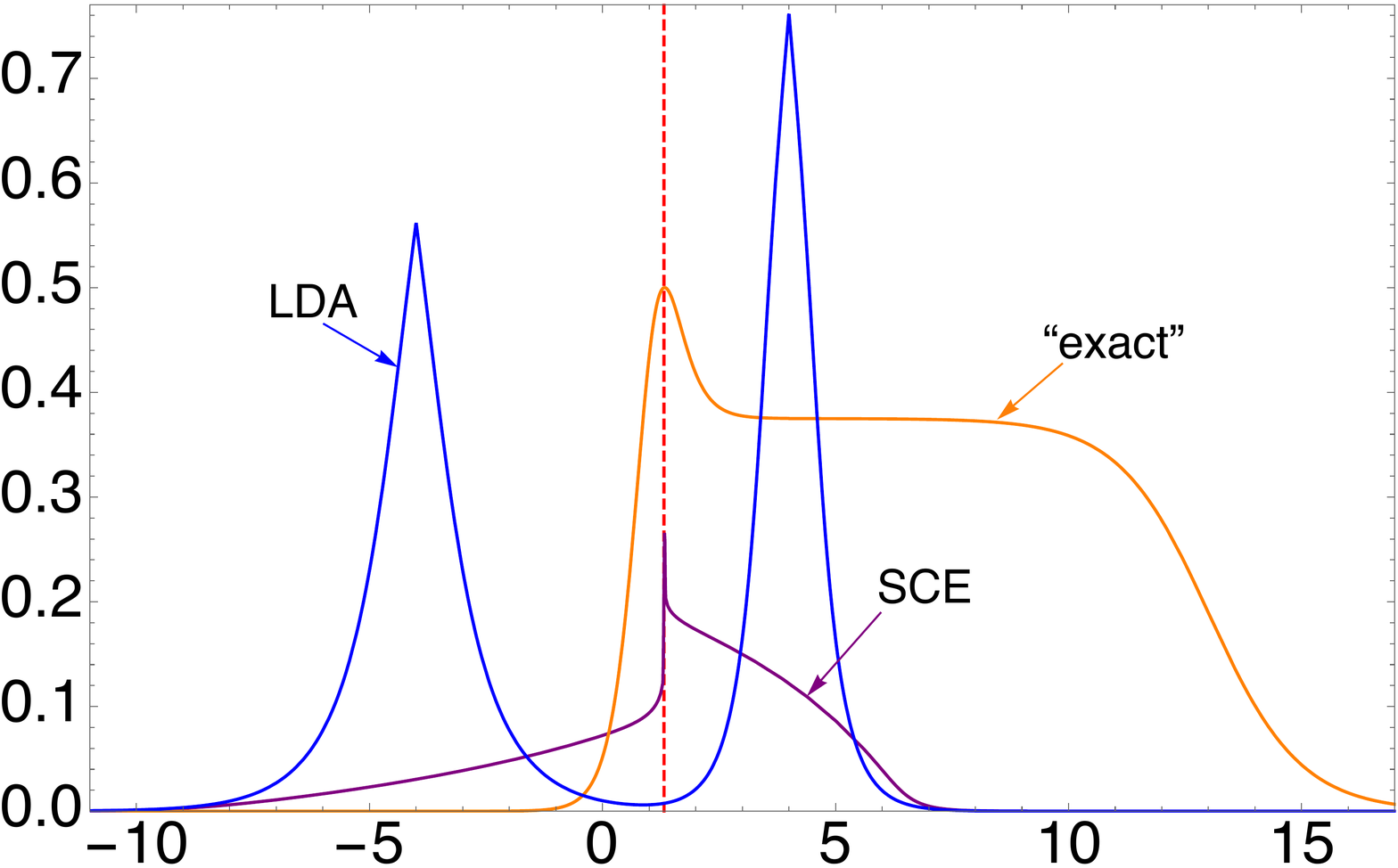}
\caption{SCE response potential compared to the ``exact'' and the LDA $\overline{v}_{resp}(x)$ for the model density in eq~\ref{rhomod} with $a=2$, $b=1$, and $R=8$. The red dashed line highlights the position where $x=a_R$.}
 \label{fgr:vsvresce8}
\end{figure}

In 1D, we have (see eq~\ref{cumulant} for comparison)
\begin{equation}
N_e(x)= \int _{-\infty}^x \rho(s) \md s
\end{equation}
and, as we have two electrons, there is only one of the ``SCE shell" borders, $a_i$, appearing in eq~\ref{comff}, 
\begin{equation}\label{aR}
a_R \, : \quad \int _{-\infty}^{a_R} \rho(x) \md x = 1.
\end{equation}
We have used the subscript ``$R$ " because the distance $a_1$ is a function of the separation between the centers of the exponentials in eq~\ref{rhomod}. Also, there is only one co-motion function that describes the position of one electron given the position $x$ of the other,
equal to\cite{Sei-PRA-99, MalMirGieWagGor-PCCP-14, ColDepDim-CJM-15}
\begin{equation}\label{1D2el-cf}
f(x)= 
\Bigg\{
\begin{array}{l}
N_e^{-1}[N_e(x) + 1] \qquad \qquad \; x < a_R \\ 
N_e^{-1}[N_e(x)-1]\qquad \qquad \;  x > a_R.
\end{array} 
\end{equation}
We have stressed in the previous section that the border of a shell that contains one electron coincides with the reference position at which one of the co-motion functions diverges. The same is true when $x\to a_R$, except that in the one-dimensional case the electron that goes to infinity has to ``reappear'' on the other side, $\lim_{x \rightarrow a_R^{\pm}} f(x) = \mp \infty $. Moreover, as we have only 2 electrons, we can use eq~\ref{vrespsce} to compute $v_{resp}^{SCE}(\br)$,
\begin{equation}
v_{resp}^{SCE} (x) = - v^{SCE}(f(x)) + v^{SCE}(a_R),
\end{equation}
which further shows that
\begin{equation}
v_{resp}^{SCE} (a_R) = v^{SCE}(a_R).
\end{equation}
In fig~\ref{fgr:vsvresce8} we show the SCE response potential compared to the ``exact'' $\overline{v}_{resp}(x)$ for the model density of eq~\ref{rhomod} at internuclear separation $R=8$, using $a=2$ and $b=1$. In the same figure, we also show the local-density approximation (LDA) CCA response potential $\overline{v}_{resp}^{LDA}(x)$ computed, as in ref~\citenum{GriMenBae-JCP-16}, via eq~\ref{eq:vbarrespfromdiff},
\begin{equation}
	\overline{v}_{resp}^{LDA}(x)=v_{xc}^{LDA}(x)-2\,\epsilon_{xc}^{LDA}(x).
	\label{eq:vrespLDA}
\end{equation}
We stress that eq~\ref{eq:vrespLDA} is the correct definition of $\overline{v}_{resp}^{LDA}(x)$, since the energy density in LDA does not have any gauge ambiguity, being given exactly in terms of the electrostatic potential associated with the CCA exchange-correlation hole of the uniform electron gas.\cite{GiuVig-BOOK-05} For the one-dimensional $\epsilon_{xc}^{LDA}$, we have used the parametrization of Casula et al.,\cite{CasSorSen-PRB-06} in which the electron-electron Coulomb interaction is renormalized at the origin,\cite{GiuVig-BOOK-05} with thickness parameter $b=0.1$. Notice that the SCE response potentials evaluated with the full Coulomb interaction $1/|x|$ or with the interaction renormalized at the origin\cite{GiuVig-BOOK-05} are indistinguishable on the scale of fig~\ref{fgr:vsvresce8}, since in the SCE limit the electron-electron distance $|x-f(x)|$ for a stretched two-electron ``molecule'' never explores the short-range part of the interaction.

The ``exact'' $\overline{v}_{resp}(x)$  has been computed by inverting the KS equation for the doubly occupied ground-state orbital $\sqrt{\rho(x)/2}$, disregarding the external potential given by attractive delta functions located at the ``nuclei'', and assuming that, for the stretched molecule, the interaction between fragments is negligible (which is asymptotically true), while the contributions coming from the Hartree potential on each fragment (the self-interaction error) are exactly canceled by the XC hole. In other words, when $R$ is large, we have $v_{Hxc}(x)\approx\overline{v}_{resp}(x)\approx v_{c,kin}(x)+v_{resp}(x)$. 

We see that, as well known, the LDA response potential completely misses the peak and the step structure of the ``exact'' $\overline{v}_{resp}(x)$, being, instead, way too repulsive on the atoms,\cite{GriMenBae-JCP-16} and following essentially the density shape. The SCE response potential, instead, even though clearly not in agreement with the ``exact'' one, shows an interesting structure located at the peak of $\overline{v}_{resp}(x)$, and also a sort of step-like feature. 

\begin{figure}
\includegraphics[scale=0.5]{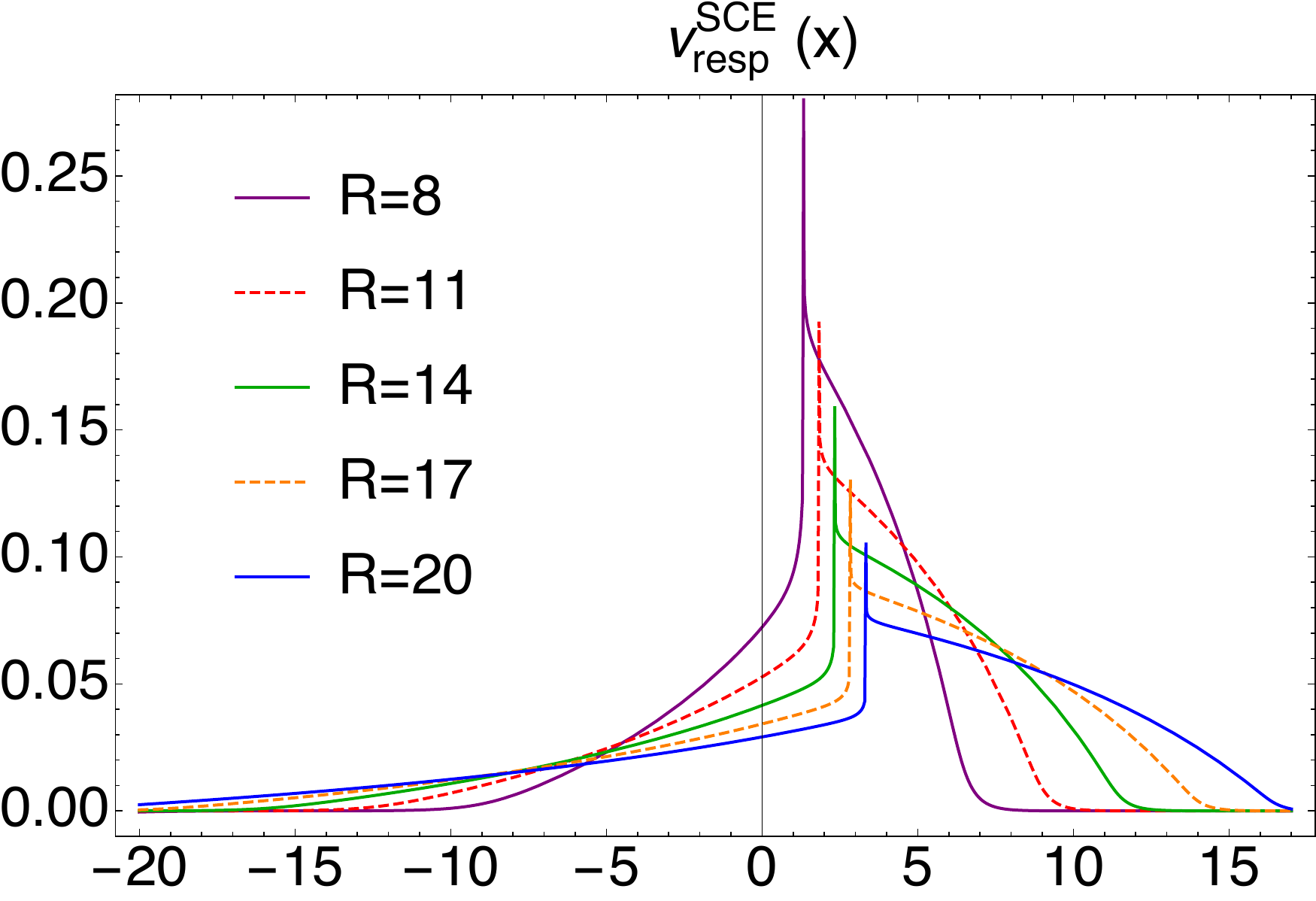}
\caption{SCE response potential for the model density in eq~\ref{rhomod} with $a=2$, $b=1$, and increasing internuclear distances, $R$.}
\label{vresce820}
\end{figure}

In fig~\ref{vresce820} we illustrate the behavior of the SCE response potential alone as the internuclear separation $R$ grows, for the same values of $a$ and $b$ of fig~\ref{fgr:vsvresce8}. We see that the SCE response potential, contrary to the exact one, does not saturate to a step height equal to the difference of the ionization potentials of the two fragments, $\Delta I_p= |I_a- I_b|$. On the contrary, $v_{resp}^{SCE}(x)$ goes (although very slowly) to zero in the dissociation limit, similarly to what happens for the midbond peak in a homodimer, as explained in refs~\citenum{MalMirGieWagGor-PCCP-14, YinBroLopVarGorLor-PRB-16}.
 This has to be expected, in view of the fact that, in the SCE limit, we are only taking into account the expectation of the Coulomb electron-electron interaction, which, when considering two distant one-electron fragments as in this case, is a vanishing contribution.\cite{YinBroLopVarGorLor-PRB-16} 
The fact that we still observe the SCE response structure for quite large $R$ values is related to the non-locality of the SCE potential and to the long-range nature of the Coulomb interaction. A kinetic contribution to SCE is clearly needed, something that is being currently investigated by looking at the next leading terms in the $\lambda\to\infty$ expansion.\cite{GorVigSei-JCTC-09, GroKooGieSeiCohMorGor-JCTC-17}

The peak structure of the SCE response potential is located at $a_R$ of eq~\ref{aR}, which is given by
\begin{equation}\label{27-29}
a_R = \frac{R}{2}\frac{(a-b)}{(a+b)} = \frac{R}{2}\frac{\left(1 - \sqrt{\frac{I_b}{I_a}}\right)}{\left(1 + \sqrt{\frac{I_b}{I_a}}\right)}.
\end{equation}
If we compare this result with the one for the location of the step in the exact KS potential, given by eqs. (27) and (29) of ref~\citenum{TemMarMai-JCTC-09}, we see that the two expressions differ by the term $\frac{1}{\sqrt{32}}\frac{\ln\frac{I_b}{I_a}}{\sqrt{I_b} + \sqrt{I_a}}$, which becomes comparatively less important as the bond is stretched. In fig~\ref{fgr:vsvresce8} we have reported the case $a=2$, $b=1$, and $R=8$, for which eq~\ref{27-29} gives $a_R = \frac{4}{3}$ and the correction term for the actual position of the step,\cite{TemMarMai-JCTC-09} which is also the position at which the kinetic peak has its maximum,  $x_{step} = x_{peak}$, gives $\frac{1}{\sqrt{32}}\frac{\ln\frac{I_b}{I_a}}{\sqrt{I_b} + \sqrt{I_a}} \simeq -0.23 $. 
The reason why, in spite of this significative correction, in fig~\ref{fgr:vsvresce8} the peak of the ``exact'' $\overline{v}_{resp}(x)$ visibly coincides with $a_R$ will be clear in the following section~\ref{careful}.

\subsection{Behaviour of the co-motion function for increasing internuclear distance}

\begin{figure}
\includegraphics[scale=0.5]{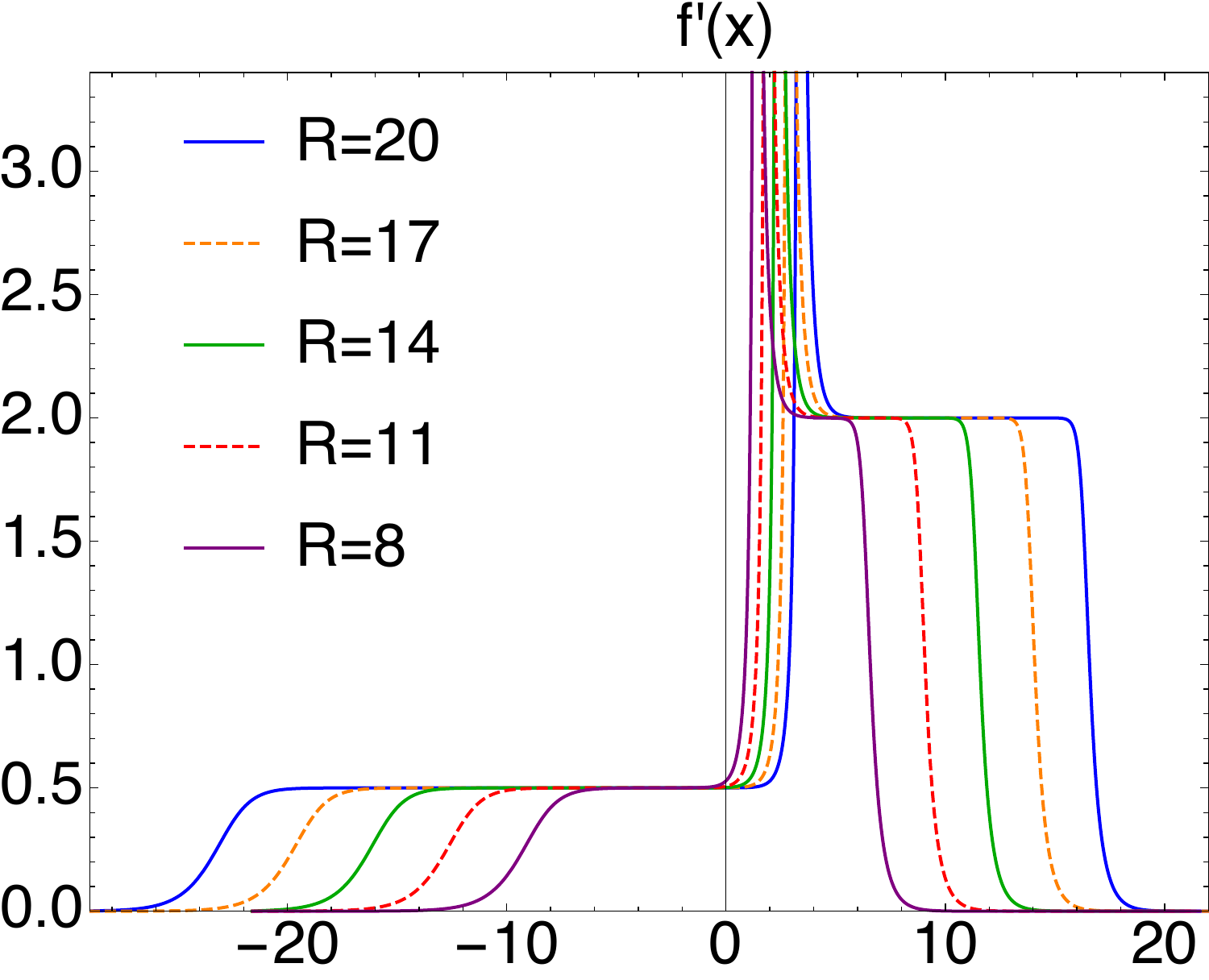}
\caption{Derivative of the co-motion function for the model density in eq~\ref{rhomod} with $a=2$, $b=1$, and increasing internuclear distances, $R$.}
\label{slope-cfR}
\end{figure}

The features of the SCE response potential can be understood by looking at how the co-motion function changes with increasing internuclear separation $R$. In the 1D two-electron case considered here, eq~\ref{differential}  becomes
\begin{equation}
f'(x)= \frac{\rho(x)}{\rho(f(x))}.
\end{equation}
For $R>>0$,  when the reference electron ($e_1$) is in the center of one the two ``atomic'' densities, e.g., at $x=-\frac{R}{2}$, the other electron ($e_2$) is in the center of the other ``atom'', $f(-\frac{R}{2}) = \frac{R}{2}$. This is a simple consequence of the fact that the overall density is normalized to two and, if the overlap in the midbond region is negligible, for symmetry reasons, the area from $-\frac{R}{2}$ to $\frac{R}{2}$ is exactly equivalent to the sum of the areas outside that range.\\
We see that after a critical internuclear distance, $R_c$, at which the overlap between the densities of the separated fragments becomes negligible, the slope of the co-motion function when $e_1$ is in $x=-\frac{R}{2}$ becomes equal to
\begin{equation}
\label{eq:fprimeStretched}
f'(x)|_{x=-\frac{R}{2}}= \frac{\rho(-\frac{R}{2})}{\rho(\frac{R}{2})}\simeq\frac{\rho_b (-\frac{R}{2})}{\rho_a(\frac{R}{2})}= \frac{b}{a} \quad R > R_c,
\end{equation}
 so that there is a region where $f'(x) = \frac{b}{a}$, and, similarly, another region where $f'(x) = \frac{a}{b}$, by interchanging $e_1$ with $e_2$.
Notice that the extension of these regions is different for the two branches of eq~\ref{1D2el-cf} and it is wider when the reference electron is around the least electronegative ``atom'' as it can be seen in fig~\ref{slope-cfR},  where we show the (numerically) exact
\begin{equation}
f'(x)=\frac{\rho_a(x- \frac{R}{2})+ \rho_b(x+ \frac{R}{2})}{\rho_a(f(x)-\frac{R}{2})+\rho_b(f(x)+\frac{R}{2})}.
\end{equation}
There, the two regions clearly appear as left and right plateaus, with their extent increasing linearly with $R$. These plateaus are the signature of molecular dissociation: they are absent at equilibrium distance, and start to appear as the overlap between the two densities is small. We see from eq~\ref{eq:fprimeStretched} that they encode information on the ratio between the ionization potentials of the two fragments.

\subsection{Careful inspection of the exact features of the KS potential for the dissociating AB molecule}\label{careful}

\begin{table} 
\small
  \caption{Some of the relevant analytic features of the analytic 1D model dimer. The table has two parts: $x< \frac{R}{2}$, and $x> \frac{R}{2}$.  In the first part  $x_{\mathrm{peak}}^{(1)}$ is the position at which the kinetic potential, $v_{c, kin} (x)$, has a maximum in between the two nuclear centers; $x_{\mathrm{step}}^{(1)}$ is the (coinciding) position at which the response potential, $v_{resp}(x)$, has an inflection point. With the subscript ``flex'' we indicate the inflection point of both the total Hartree-XC potential and the kinetic potential; they are distinguished via an additional subscript, respectively ``Hxc'' and ``k''.  Finally, $x_{eq} $ is used to label the x-value at which  $v_{c, kin} (x)$ and $v_{resp}(x)$ crosses.  In the second part, the analogous quantities to the ones just explained, appearing in this case somewhere far from the midbond on the side of the more electronegative fragment, are listed. For example, $x_{\mathrm{peak}}^{(2)}$ is the second maximum of the kinetic potential, eq~\ref{tab:eq2} (top-right entry of the second part), which also coincides with the second inflection point of the response potential as argued in the main text.}
   \label{tbl:list}
  \begin{ruledtabular}
  \begin{adjustbox}{width=0.35\textwidth}
\small
  \begin{tabular*}{0.5\textwidth}{@{\extracolsep{\fill}}lllr}
   $a\geqslant b$ &    
$\begin{array}{l} \phi_a(x)=\sqrt{\frac{a}{2}}e^{-\frac{a}{2}|x-\frac{R}{2}|}\\
\phi_b(x)=\sqrt{\frac{b}{2}}e^{-\frac{b}{2}|x+\frac{R}{2}|}
\end{array}$  &      $\begin{array}{l}\rho(x)= |\phi_a(x)|^2+ |\phi_b(x)|^2\\  I_{\alpha} = \frac{\alpha^2}{8}; \quad \alpha = a, b \end{array}  $ &$\quad$\\
 
      \hline
  & & $\quad$\\  
   $x < \frac{R}{2}$ &    $\begin{scriptsize} \frac{\md v_{c, kin}(x)}{\md x}|_{x_{\mathrm{peak}}^{(1)}}=\frac{\md^2 v_{resp}(x)}{\md x^2}|_{x_{\mathrm{step}}^{(1)}}   \end{scriptsize} $ &  
$
  x_{\mathrm{peak}}^{(1)}=\frac{(a-b)R + 2 \ln{\frac{b}{a}}}{2 (a+b)}
$ \tagarray\label{tab:eq1}&$\quad$\\
&   & $v_{c, kin}(x_{\mathrm{peak}}^{(1)}) = \frac{1}{8}\left(\frac{a+b}{2}\right)^2$  &$\quad$ \\
&   & $v_{c, kin}(x_{\mathrm{peak}}^{(1)}) = \frac{1}{8}\left(\frac{a+b}{2}\right)^2$  & $\quad$\\

&   & $v_{resp}(x_{\mathrm{step}}^{(1)})= \frac{1}{2}\frac{a^2-b^2}{8}$ &$\quad$ \\
 &    & $v_{Hxc}(x_{\mathrm{peak}}^{(1)}) = \frac{1}{32}(3 a - b) (a+b)$ & $\quad$\\
  &     $\frac{\md^2 v_{c,kin} (x)}{\md x^2}|_{x_{\mathrm{flex},k}^{(1)}}=0$ & $ x_{\mathrm{flex},k^{(1)}}=\frac{(a-b)R - 2\ln{\frac{2a + \sqrt{3}a}{b}}}{2 (a+b)}$& $\quad$\\
 &     $\frac{\md^2 v_{c,kin} (x)}{\md x^2}|_{x_{\mathrm{flex},k}^{(2)}}=0$ & $ x_{\mathrm{flex},k^{(2)}}=\frac{(a-b)R - 2\ln{\frac{2a - \sqrt{3}a}{b}}}{2 (a+b)}$& $\quad$\\  
  
 &     $\frac{\md v_{Hxc} (x)}{\md x}|_{x_{\mathrm{peak},Hxc}=0}$ & $ x_{\mathrm{peak},Hxc}=\frac{(a-b)}{(a+b)}\frac{R}{2} = a_R$ &$\quad$ \\
&       $\frac{\md^2 v_{Hxc} (x)}{\md x^2}|_{x_{\mathrm{flex},xc}^{(1)}}=0$&$x_{\mathrm{flex},xc}^{(1)}=\frac{(a-b) R - 2 \ln{\frac{(a + b + \sqrt{a^2 + ab + b^2})}{b}}}{2 (a+b)}$& $\quad$\\
&$ $&$ $&$\quad$ \\
 &      $\frac{\md^2 v_{Hxc} (x)}{\md x^2}|_{x_{\mathrm{flex},xc}^{(2)}}=0$&$x_{\mathrm{flex},xc}^{(2)}=\frac{(a-b) R - 2 \ln{\frac{(a + b - \sqrt{a^2 + ab + b^2})}{b}}}{2 (a+b)}$& $\quad$\\
       
& $ v_{c,kin}(x_{eq})=v_{resp}(x_{eq})$ & $x_{eq}= \frac{(a-b)R + 2 \ln{2} + 2 \ln{\frac{b^2}{a (b-a)}}}{2 (a+b)}$& $\quad$\\
\hline
  & & & $\quad$\\
$x > \frac{R}{2} $ &   $ \begin{scriptsize} \frac{\md v_{c, kin}(x)}{\md x}|_{x_{\mathrm{peak}}^{(2)}}=\frac{\md^2 v_{resp}(x)}{\md x^2}|_{x_{\mathrm{step}}^{(2)}}   \end{scriptsize} $ &   $
x_{\mathrm{peak}}^{(2)}=\frac{(a+b)R - 2 \ln{\frac{b}{a}}}{2(a-b)} $ \tagarray\label{tab:eq2}& $\quad$\\
 &    &   $v_{c, kin}(x_{\mathrm{peak}}^{(2)}) = \frac{1}{8}\left(\frac{a-b}{2}\right)^2$& $\quad$ \\

&      & $v_{resp}(x_{\mathrm{step}}^{(2)})= v_{resp}(x_{\mathrm{step}}^{(1)})$& $\quad$\\
&    &   $v_{Hxc}(x_{\mathrm{peak}}^{(2)}) = \frac{1}{32}(3 a + b) (a - b)$ & \\
 &     $\frac{\md^2 v_{c,kin} (x)}{\md x^2}|_{x_{\mathrm{flex},k}^{(3)}}=0 $ & $ x_{\mathrm{flex},k^{(3)}}=\frac{(a+b) R - 2 \ln{\frac{(2b + \sqrt{3}b)}{a}}}{2 (a-b)}$& $\quad$\\
 &     $\frac{\md^2 v_{c,kin} (x)}{\md x^2}|_{x_{\mathrm{flex},k}^{(4)}}=0$ & $ x_{\mathrm{flex},k^{(4)}}=\frac{(a+b) R - 2 \ln{\frac{(2b - \sqrt{3}b)}{a}}}{2 (a-b)}$& $\quad$\\  
  
&     $ \frac{\md^2 v_{Hxc} (x)}{\md x^2}|_{x_{\mathrm{flex},xc}^{(3)}}=0$ & $x_{\mathrm{flex},xc}^{(3)}=\frac{(a+b) R + 2 \ln{\frac{(a - b + \sqrt{a^2 - ab + b^2})}{b}}}{2 (a - b)}$& $\quad$\\
  \end{tabular*}
  \end{adjustbox}
  \end{ruledtabular}
\end{table}

\begin{figure}
\includegraphics[scale=0.3]{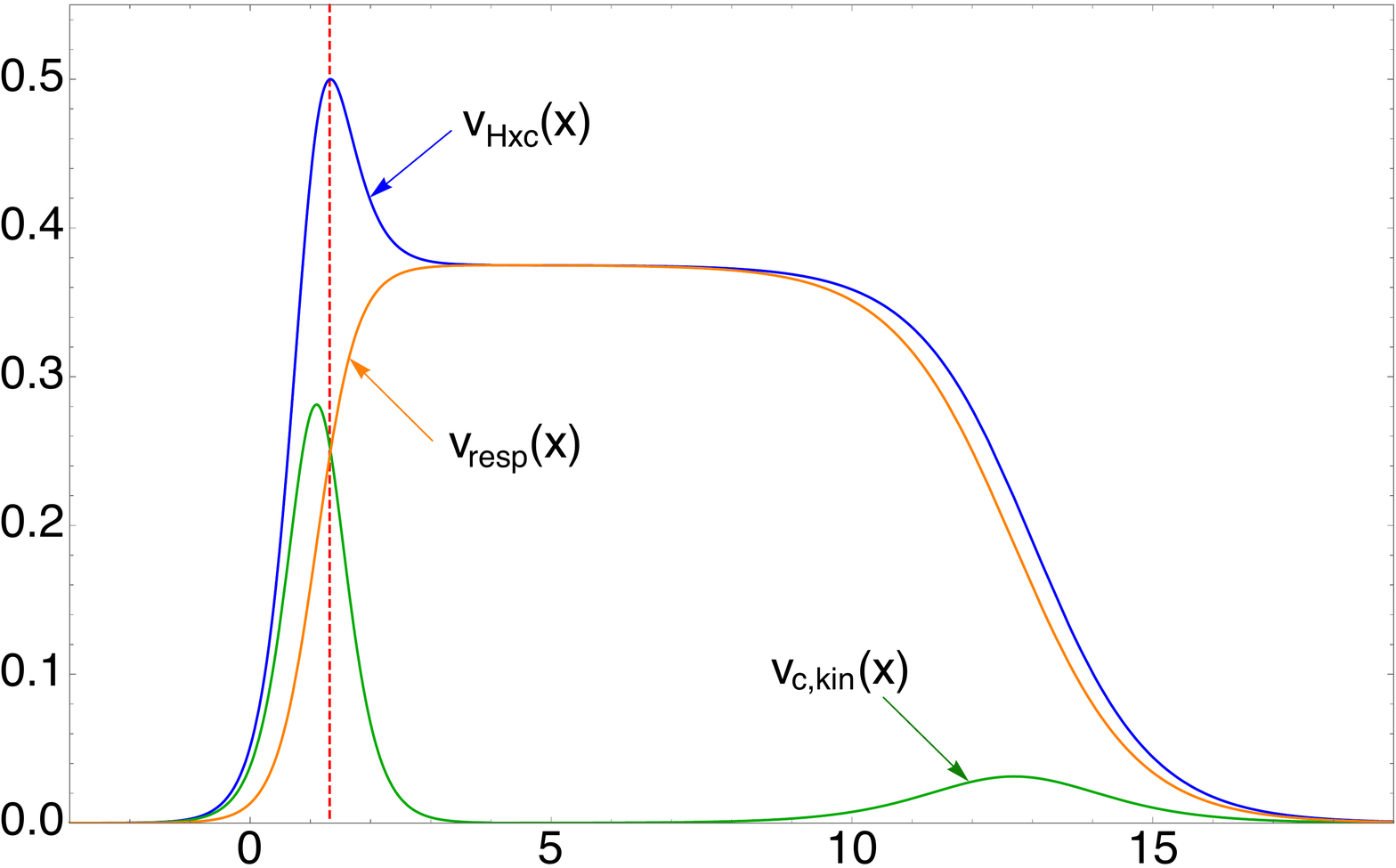}
\caption{Hartree-XC potential, $v_{Hxc}(x)$, and its contributions $ v_{c,kin}(x)$ and $v_{resp}(x)$ for $a=2, b=1, R=8$. The red dashed line highlights the position where $x=a_R$. }
\label{vsvrevck}
\end{figure}

The model density $\rho(x)$ described of eq~\ref{rhomod} corresponds to an asymptotic simplification of different models that appeared in the literature to study the KS potential in the dimer dissociation limit.\cite{TemMarMai-JCTC-09, HelTokRub-JCP-09, HodRamGod-PRB-16, BenPro-PRA-16}
Here we review in detail the properties of the KS potential and the two single contributions that can be extracted from this model, $v_{c, kin}(x)$ (eq~\ref{vckin}) and $v_{resp}(x)$ (eq~\ref{vresp} or \ref{respN-1}), also showing that a second peak in the kinetic potential appears on the side of the most electronegative ``atom'', a feature that seemed to have been overlooked in previous studies.
In order to study the dissociation regime we use the Heitler-London wavefunction:
\begin{small}
\begin{equation}\label{HLWF}
\Psi_{HL}(x_1, x_2)=\frac{1}{\sqrt{2 (1 + S_{AB})}} \left( \phi_a(x_1) \phi_b(x_2) + \phi_b(x_1)\phi_a(x_2) \right),
\end{equation}
\end{small}
where $S_{AB}= \int \phi_a(x)\phi_b \md x$, and $\phi_{a,b}= \sqrt{\frac{a}{2}}e^{-\frac{a,b}{2} |x\pm\frac{R}{2}|}$.
To compute the kinetic potential, in the dissociation limit,  we can use eq~\ref{vkin} and the conditional amplitude coming from the Heitler-London wavefunction considering $S_{AB}=0$, which yields the well-known expression \cite{TemMarMai-JCTC-09, HelTokRub-JCP-09}
\begin{align}\label{vckinmod}
v_{c, kin}(x)&= \frac{1}{2}\int |\frac{d}{dx} \Phi_{HL}(x_2 | x)|^2 \md x_2   \nonumber \\ &=\frac{1}{2} \frac{\left(\phi_b(x)\phi_a'(x) - \phi_a(x)\phi_b'(x)\right)^2}{\left( \phi_a(x)^2 + \phi_b(x)^2  \right)^2},
\end{align}
where we have used the fact that $v_{kin}(x)=v_{c, kin}(x)$ as the kinetic KS potential is zero for a closed-shell two-electron system. Analogously, $v_{resp}(x)$ can be obtained from $v_{N-1}$ of eq~\ref{N-1},
\begin{align}\label{vrespmod}
v_{resp}(x)&= \frac{1}{2} \int |\frac{d}{dx_2} \Phi_{HL}(x_2 | x)|^2 \md x_2  +&  \nonumber \\
& + \int v_{ext}^{mod}(x_2) |\Phi_{HL}(x_2 | x)|^2 \md x_2  - E^{N-1}= \nonumber \\
&= - \frac{1}{\rho(x)} \left( \frac{a^2}{8} \phi_b(x)^2 + \frac{b^2}{8} \phi_a(x)^2 \right) + \frac{a^2}{8},
\end{align}
where  $v_{ext}^{mod}(x) = -\frac{a}{2} \delta \left( x - \frac{R}{2}\right) -\frac{b}{2} \delta\left(x + \frac{R}{2}\right)$ and $E^{N-1} = - \frac{a^2}{8}$. 
Comparing these two contributions with the KS potential obtained from the density by inversion (subtracting the external potential due to the attractive delta peaks at the ``nuclear'' positions), we have in this limit, as already discussed,
\begin{equation}
 v_{Hxc}(x)\approx \overline{v}_{resp}(x) \approx v_{c,kin}(x) + v_{resp}(x),
\label{vrespendo}
\end{equation}
since  $v_{cond}(x)$ goes to zero when the fragments are very far from each other.
In fig.~\ref{vsvrevck} we show the potential obtained from the inversion of the KS equation with its two components $v_{c,kin}(x)$ and $v_{resp}(x)$.

For this simple model, we have exact expressions regarding each component of the potential and their maxima, inflection points, and so forth. Some of these relevant analytic expressions are listed in table \ref{tbl:list}.
By looking at the table, one sees, for example, that the peak of the total Hartree-XC potential is not located where the peak of the kinetic correlation builds up.
In particular the maximum of the Hartree-XC potential is found at 
\begin{equation}\label{xspeak}
x_{peak, Hxc} = \frac{R}{2}\frac{(a-b)}{(a+b)},
\end{equation}
 which is exactly $a_R$ (see eq~\ref{27-29} and compare also fig.~\ref{fgr:vsvresce8}).
Thus, the Hartree-XC potential potential reaches its maximum when the density integrates to one electron (or the correct integer number of electrons in a general two fragments case) because this is where the two fragments must be detached from one another. From a different perspective, this is a manifestation that the response and the kinetic correlation contributions in the dissociation limit are not independent and that their sum can be sometimes more meaningful than the separate contributions.
Also, by playing around with the expressions in table \ref{tbl:list}, one realises that there can be misleading coincidental features. For example, the last entry of the first section of the table, which is the analytic expression for the distance at which the kinetic correlation potential and the response potential equate, $ x_{eq} $,  is such that the two contributions $v_{c, kin}(x)$ and $v_{resp}(x)$ crosses exactly at $a_R$  if $a=2$ and $b=1$ like in fig.~\ref{vsvrevck}, but this is not a general feature.  Similarly if we choose $a=\frac{5}{3} b$  then the height of the kinetic peak becomes equal to the height of the step and so on. \\
Note here that the features listed in the table are obtained for the zero-overlap case, $S_{AB} = 0$, in  eq~\ref{HLWF}. Nonetheless, they should become asymptotically
 exact in the dissociation limit.
 
Another feature that came to our attention and that -- to the best of our knowledge -- has not been discussed before, is the fact that the kinetic correlation potential has a second peak on the side of the most electronegative atom.
This second maximum is located where the second inflection point of the response potential is, see fig.~\ref{vsvrevck} and eq~\ref{tab:eq2} in tab.~\ref{tbl:list}.
To understand the appearance of the second peak, we can identify two regimes, A and B, by the leading exponential coefficient: for example, in our case, in the region starting from $-\infty$ the density of the fragment with the smallest coefficient, $\rho_b(x)$ is larger than the other, $\rho_a(x)$; approaching the A center there is a point in which $\rho_a(x)$ becomes larger than the other density. 
This transition between regimes determines both the kinetic peak and the response step.
 In particular the distance, $x^{(1)}$ at which the orbitals, $\phi_i$ (or the fragment densities, which are simply their square) equate
  \begin{equation}\label{X1}
\phi_a(x^{(1)}) = \phi_b(x^{(1)}),
\end{equation} 
 is found to coincide with that of eq~\ref{tab:eq1}, i.e. the maximum of the first kinetic peak as well as of the flex coming from the building up of the response potential step, $x^{(1)} = x_{\mathrm{peak}}^{(1)} = x_{\mathrm{step}}^{(1)}$.
 Note also that this distance is always somewhere in between the two centers of the fragments, $-\frac{R}{2} < x_{\mathrm{peak}}^{(1)} < \frac{R}{2}$.\\
Nonetheless, since $\rho_b(x)$ is asymptotically dominating, by going further in the direction of $+ \infty$, the `B regime' is to be encountered again 
and the two fragment densities, though both very small in magnitude, will be equal again, at some point, $x^{(2)}$
\begin{equation}\label{X2}
\phi_a(x^{(2)}) = \phi_b(x^{(2)}).
\end{equation} 
At this distance also another kinetic peak is appearing as well as another flex coming from the exhaustion of the response potential step or, in short, $x^{(2)} = x_{\mathrm{peak}}^{(2)} = x_{\mathrm{step}}^{(2)}$. This is in agreement with the observation in the work of Baerends and coworkers that steps in the response potential and peaks in the kinetic correlation potential are always related.\cite{BuiBaeSni-PRA-89,GriLeeBae-JCP-94}

\section{Conclusions}
In the present work we have generalised the concept of effective and response potentials, as well as of conditional amplitude, for any $\lambda$-value, and derived the modulus squared of this latter in the $\lambda\to\infty$ (SCE) limit. A consistent definition of the response potential in the SCE limit arises from our treatment.
In the simple 1D model of a dissociating molecule  (eq~\ref{rhomod}),  it is found that interesting similarities between dissociation features of the exchange-correlation potential and SCE features, such as the behaviour of the co-motion function for increasing internuclear distance or the structure of the SCE response potential itself,  can be established.
For example, in the dissociation regime, the slope of the co-motion function is determined by the ratio between the ionization potentials of the fragments (compare fig~\ref{slope-cfR}), whereby the step height of the exchange-correlation potential is determined by their difference.
In addition, the co-motion function confers to the SCE response potential an asymmetric structure which indicates on which side of the system the more electronegative fragment is located.

Further analyzing the different components of the exchange-correlation potential that are relevant in the dissociation limit, namely $v_{resp}$ and $v_{c,kin}$, or $\overline{v}_{resp}$, we have identified  the presence of a second peak of lower intensity in the kinetic correlation potential on the side of the more electronegative atom and, by comparison, we have observed that the peak of the coupling-constant averaged response potential asymptotically coincides with that of the SCE response potential itself.
Our work, together with a very recent and promising study,\cite{VucGor-JPCL-17} shows that the SCE framework encodes more than few pieces of information on the physical system, and that useful guidelines in the design of highly non-local density functional approximations (based on integrals of the density) can fruitfully be drawn from it.
A step further in this direction will be to study exact properties of the kinetic potential that appears as the next leading term ($\sim \lambda^{-1/2}$) in the expansion of the adiabatic connection integrand in the $\lambda \rightarrow \infty$ limit,\cite{GorVigSei-JCTC-09} as well as spin effects that have been shown\cite{GroKooGieSeiCohMorGor-JCTC-17} to enter at orders $\sim e^{-\sqrt{\lambda}}$.

We have also reported, for some small systems (He series, Be, Ne, and H$_2$), the response potential coupling-constant averaged along the adiabatic connection; the study of this different response potential complements that of the response potential at full coupling strength and could provide other hints for the construction of approximate XC functionals, especially of a new generation of DFAs based on local quantities along the adiabatic connection.\cite{VucIroSavTeaGor-JCTC-16, BahZhoErn-JCP-16, VucIroWagTeaGor-PCCP-17}

\section*{Acknowledgements}
We thank T.J.P. Irons and A.M. Teale for the coupling-constant averaged energy densities of sec~\ref{physical}, and E.J. Baerends for insightful discussions.
Financial support was provided by the European Research Council under H2020/ERC Consolidator Grant corr-DFT [Grant Number 648932]. 

 \appendix
 \section{Redundancy of the permutations}\label{app_perm}
In order to account for the indistinguishability among electrons the modulus squared of the SCE wavefunction has been usually expressed as (see for example eq (14) in ref~\citenum{MirSeiGor-JCTC-12})
\begin{equation} \label{starting}
\vert\Psi_{SCE}(\textbf{r}_1,\cdots,\textbf{r}_N)\vert^2=\frac{1}{N!}\sum_{\wp =1}^{N!}\int d\textbf{s}\frac{\rho(\textbf{s})}{N}\prod_{i=1}^N \delta(\textbf{r}_i -\textbf{f}_{\wp (i)}(\textbf{s})).
\end{equation}
We want to show here that, by virtue of the two basic properties of the co-motion functions, eqs~\ref{differential} and \ref{eq:groupprop}, all the permutations contribute in the same way to the potentials computed from the SCE conditional amplitude of eq~\ref{CA}), and thus the use of eq~\ref{starting} is formally equivalent to eq~\ref{PsiSCE} in this context. 
If we perform the integration over $\textbf{s}$ for all the permutations we can rewrite eq~\ref{starting} as:
\begin{small}
\begin{align} 
&\vert\Psi_{SCE}(\textbf{r}_1,\cdots,\textbf{r}_N)\vert^2 =\nonumber\\
& = \frac{1}{N!}  \sum_{\wp =1}^{(N-1)!}  \left( \frac{\rho(\br_1)}{N} \prod_{i=2}^N  \delta(\br_i -\fv_{\wp (i)}(\br_1)) + \right. \nonumber \\
&  + \frac{\rho(\br_2)}{N}\prod_{i=1,3,\cdots, N}  \delta(\br_i -\fv_{\wp (i)}(\br_2)) + \nonumber \\
& \left. \cdots + \frac{\rho(\br_N)}{N}\prod_{i=1} ^{(N-1)} \delta(\br_i -\fv_{\wp (i)}(\br_N) ) \right)  \label{continuing}
\end{align}
\end{small}
Now we want to show that each of the $N!$ terms inside brackets in eq~\ref{continuing} will have the same contribution to the potentials computed from the conditional amplitude. Since the variables $i=2,...N$ are always integrated out in a symmetric way in the computation of the effective potentials, all what we need to show is that all the terms have the prefactor $\rho(\br_1)$ in front. We perform the explicit computation for the 3-electron case, from which it becomes clear that the reasoning applies also to the general $N$-electron case.
For $N=3$ we have $\wp =1\cdots 6$, so that the wavefunction reads
\begin{footnotesize}
\begin{align*}
&\vert\Psi_{SCE}(\br_1, \br_2, \br_3)\vert^2 =\\
&\frac{1}{6} \! \Big[ \! \underbrace{\frac{\rho(\br_1)}{3}}_{\wp=2}\Big(\! \delta(\br_2 -\fv_{2}(\br_1)) \delta(\br_3 -\fv_{3}(\br_1)) +  \underbrace{\delta(\br_2 -\fv_{3}(\br_1)) \delta(\br_3 -\fv_{2}(\br_1))}_{\wp=2} \!\Big)\\
& + \underbrace{\frac{\rho(\br_2)}{3}  \Big( \delta(\br_1 -\fv_{2}(\br_2))\delta(\br_3 -\fv_{3}(\br_2))}_{\wp = 4} + \delta(\br_1 -\fv_{3}(\br_2))\delta(\br_3 -\fv_{2}(\br_2)) \Big)\\
& + \frac{\rho(\br_3)}{3}  \Big(\delta(\br_1 -\fv_{2}(\br_3))\delta(\br_2 -\fv_{3}(\br_3) + \delta(\br_1 -\fv_{3}(\br_3))\delta(\br_2 -\fv_{2}(\br_3) \Big)  \Big]
\end{align*}
\end{footnotesize}
We now consider one permutation, e.g.~the underlined $\wp = 4$ term
\begin{equation*}
\frac{\rho(\br_2)}{3} \delta(\br_1 -\fv_{2}(\br_2)) \delta(\br_3 -\fv_{3}(\br_2),
\end{equation*} 
in the following we are going to show that this term is equivalent to the  $\wp = 2$ term (also highlighted for the purpose).
\begin{enumerate}
\item Using the basic property of change of variables in the delta function on $\delta(\br_1 -\fv_{2}(\br_2)) $, we can rewrite this permutation as
\begin{small}
\begin{equation} \label{JD}
\frac{\rho(\fv_2^{-1}(\br_1))}{3 \:\mathrm{det}\big(\partial_{f_{2,\alpha}^{-1}} f_{2,\beta} (\fv_2 ^{-1} (\br_1))\big)} \delta\big(\br_2 - \fv_2^{-1}(\br_1)\big) \delta\big(\br_3 - \fv_3(\fv_2^{-1}(\br_1))\big),
\end{equation} 
\end{small}
where the indices $\alpha,\beta=x,y,z$, and $\mathrm{det}\big(\partial_{\alpha} g_{\beta} (\br)\big)$ denotes the Jacobian of the transformation ${\bf g}(\br)$.
\item Using the property of the inverse function we can rewrite this term as 
\begin{equation}\label{JD2}
\frac{\rho(\fv_2^{-1}(\br_1))}{3} \mathrm{det}\big(\partial_{\alpha} f_{2,\beta} ^{-1} (\br_1)\big) \delta\big(\br_2 - \fv_2^{-1}(\br_1)\big) \delta\big(\br_3 - \fv_3(\fv_2^{-1}(\br_1))\big).
\end{equation}
\item Finally, using eqs~\ref{differential} and ~\ref{eq:groupprop}, which imply that the inverse of a co-motion function is another co-motion function, the term (\ref{JD2}) transforms into
\begin{equation}
\frac{\rho(\br_1)}{3} \delta (\br_2-\fv_3(\br_1)) \delta (\br_3-\fv_2(\br_1)),
\end{equation}
which gets the correct prefactor $\rho(\br_1)$ in front, and can also be recognised as permutation $\wp=2$.
\end{enumerate}
The same reasoning in three steps is applicable to all the terms of a general $N$-electron case.

\section{Exchange response potential for N=2 and data validation}\label{app_datavalidat}
It is common use in DFT to separate the exchange and correlation contributions in potentials and energy expressions.
Analogously to the total XC potential, the exchange potential is defined as the functional derivative of the exchange energy, which is in turn defined as
\begin{equation}
E_x[\rho] = \langle \Psi_s(1, \dots, N)|\hat{V}_{ee}| \Psi_s(1, \dots, N)\rangle - U[\rho].
\end{equation}
\begin{figure}
\includegraphics[scale=0.45]{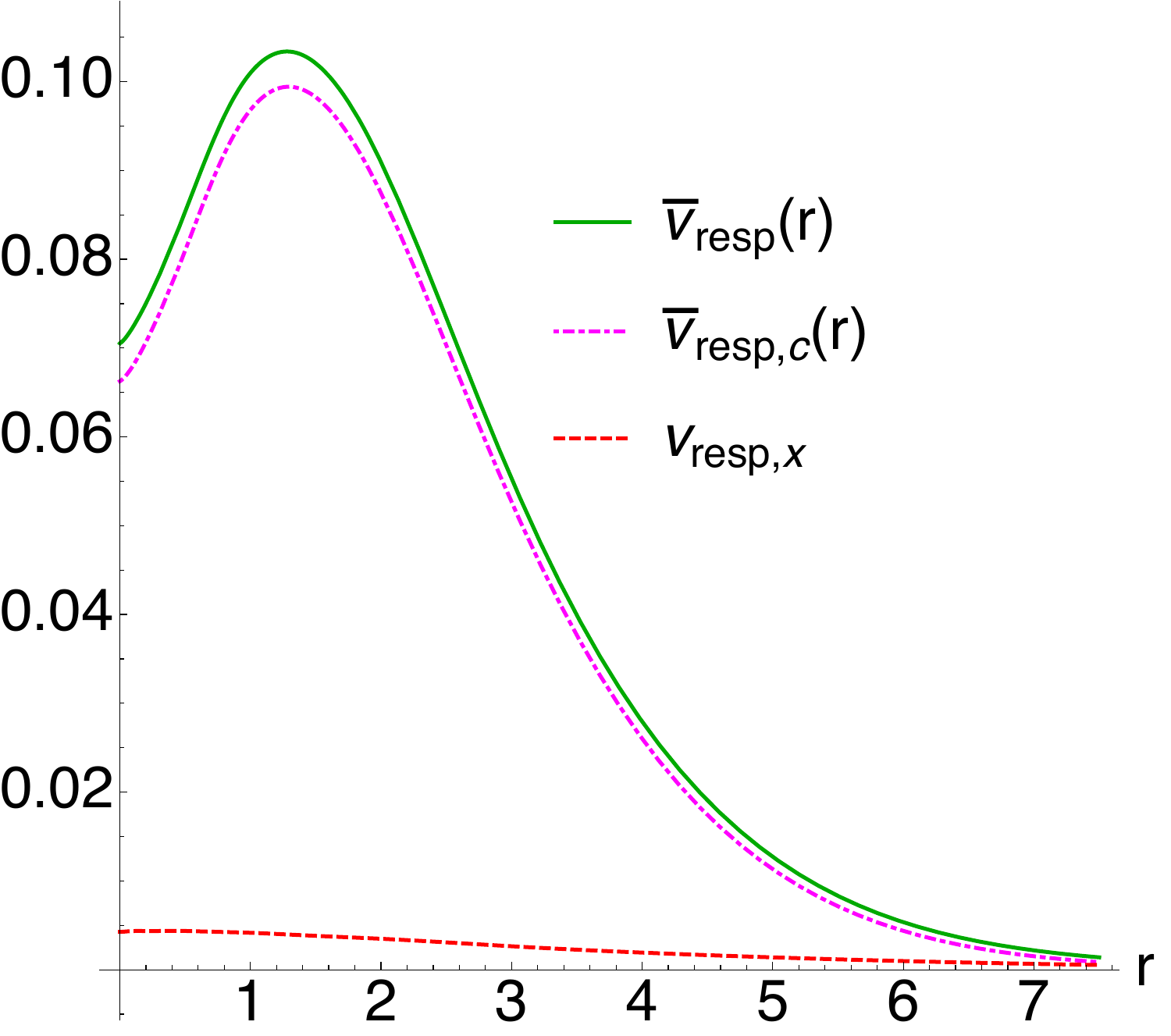}
\caption{Comparison between $\overline{v}_{resp}(r)$ and $\overline{v}_{resp, c}(\br)$ for the H$^-$ atom in order to estimate the error coming from numerics and the use of different sources for $v_{xc}(\br)$ and $\overline{w}(\br)$. }
\label{validHm}
\end{figure}
\begin{figure}
\includegraphics[scale=0.45]{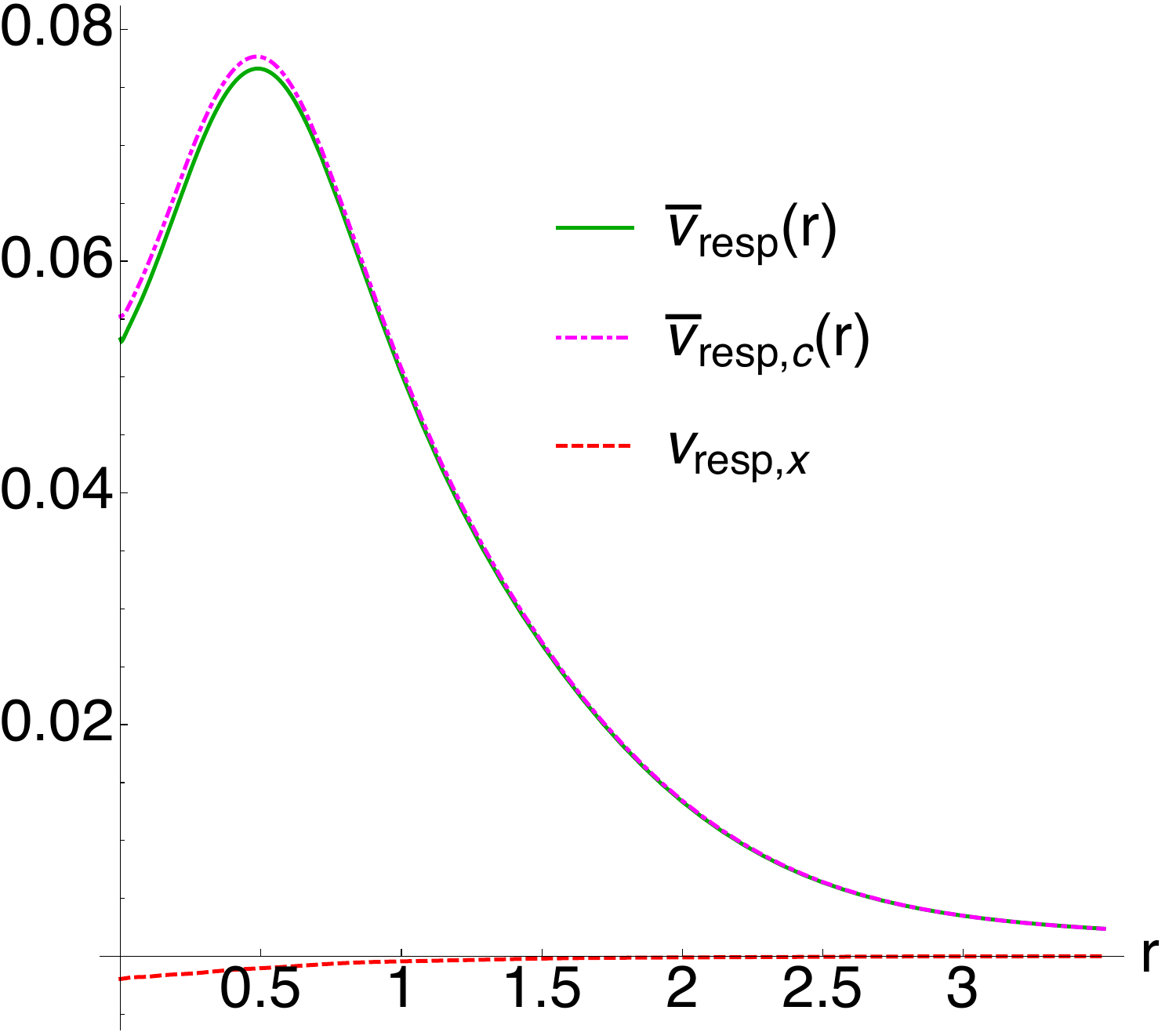}
\caption{Comparison between $\overline{v}_{resp}(r)$ and $\overline{v}_{resp, c}(\br)$ for the He atom in order to estimate the error coming from numerics and the use of different sources for $v_{xc}(\br)$ and $\overline{w}(\br)$.}
\label{validHe}
\end{figure}
For a two-electron closed-shell system we have
\begin{equation}\label{known}
E_x[\rho]=- \frac{1}{4} \int \frac{\rho(\br) \rho(\br')}{|\br -\br'|},
\end{equation}
which implies $v_{resp, x}(\br) = 0$.
In sec~\ref{physical} we have shown the CCA response potential for some atoms combining quantities coming from different sources (see eq~\ref{vrespbarprac}); namely refs~\citenum{MirUmrMorGor-JCP-14, UmrGon-PRA-94, FilGonUmr-INC-96} for the XC potentials (or their separate contributions), and refs~\citenum{IroTea-MP-15,VucIroSavTeaGor-JCTC-16} for the CCA energy densities. In the case of the H$_2$ molecule, instead, both the total XC potential and the CCA energy density used are from the latter source.

In order to give a feeling of how our results could be affected by computational inaccuracies we show in fig.~\ref{validHm} and \ref{validHe}, the difference $v_{resp, x}(\br) = v_x(\br) - 2 w_0(\br)$, together with the total $\overline{v}_{resp}(\br)$ and $\overline{v}_{resp, c}(\br) = \overline{v}_{resp}(\br) -v_{resp, x}(\br) $. The fact that the first quantity is not exactly zero and the last two are slightly different gives an idea of the numerical errors we have. As it can be noticed, the difference is between $1\div10 \%$ of the quantity of interest, $\overline{v}_{resp}(\br)$, and the discussion in sec~\ref{physres} is not affected by this error range. 

 \bibliographystyle{unsrt}

\end{document}